\documentclass[10pt,a4paper]{article}
\usepackage{amssymb}
\newcommand{\abs}[1]{\left|#1\right|}
\newcommand{\ket}[1]{\left|#1\right\rangle}
\newcommand{\bra}[1]{\left\langle#1\right|}
\newcommand{\PDM}{\opform{\rho}}
\newcommand{\opform}[1]{\widehat{#1}}
\def\kbar{{\mathchar'26\mkern-9muk}}
\newcommand{\expect}[1]{\left\langle#1\right\rangle}
\newcommand{\half}{\frac{1}{2}}

\begin{document}
\begin{center}
\textbf{\large{A Numerical Investigation of the Effects of
Classical Phase Space Structure on a Quantum System}}\\[5mm]

G. Ball, K. Vant and N. Christensen\footnote{Electronic address:
n.christensen@auckland.ac.nz}\\

\emph{Department of Physics, University of Auckland, Private Bag
92019 Auckland, New Zealand}

\date{\today}
\end{center}

\begin{abstract}
We present a detailed numerical study of a chaotic classical
system and its quantum counterpart. The system is a special case
of a kicked rotor and for certain parameter values possesses
cantori dividing chaotic regions of the classical phase space. We
investigate the diffusion of particles through a cantorus;
classical diffusion is observed but quantum diffusion is only
significant when the classical phase space area escaping through
the cantorus per kicking period greatly exceeds Planck's constant.
A quantum analysis confirms that the cantori act as barriers. We
numerically estimate the classical phase space flux through the
cantorus per kick and relate this quantity to the behaviour of the
quantum system. We introduce decoherence via environmental
interactions with the quantum system and observe the subsequent
increase in the transport of quantum particles through the
boundary.
\end{abstract}

\section{Introduction}
Quantum chaos is a relatively new field, comprising the study of
the quantized versions of systems which are classically chaotic.
The current consensus is that quantum systems do not show
sensitive dependence on initial conditions in the same way as
classical systems. This would appear to disqualify them from being
described as chaotic. However, classical chaos is also apparent in
other aspects of a system's evolution, the quantum analogies of
which are of great interest.  In particular, these include
Kolmogorov-Arnold-Moser (KAM) tori and cantori, the study of which
is the focus of this paper.  Our group has previously published
experimental studies \cite{vant,ball,christensen} of a particular,
classically chaotic system containing KAM tori and cantori, which
is realized experimentally in the quantum regime, and simulated
both classically and quantum mechanically.  This paper
concentrates on the detailed results of our computer simulations.

KAM tori and cantori in a classical phase space are predicted to
influence the corresponding quantum system. An unbroken KAM
boundary will prohibit classical diffusion through it, while
tunneling across the barrier is possible in a quantum system. When
interaction terms in the perturbing Hamiltonian are sufficiently
large as to break up the boundary and create a cantorus or
\emph{turnstile}, classical particles will quickly diffuse through
that cantorus but the quantum wavefunction will be inhibited
\cite{geisel86,geisel89,Brown}. A heuristic model proposes that
with the presence of a perturbing Hamiltonian, quantum diffusion
is constrained when the classical phase space area escaping
through the cantorus each period is $\sim \hbar$
\cite{Brown,MacKay}. Even though the barrier has been broken, the
quantum wave function still appears to tunnel through the
cantorus. 

The link between the quantum domain and the familiar classical
world remains a hotly debated topic. The quantum-classical
correspondence (QCC) principle requires that quantum mechanics
contains the classical macroscopic limit. A promising approach to
the question of quantum-classical correspondence is the study of
\emph{decoherence} --- the analysis of the effect of coupling to
the environment, which inevitably occurs in a real system, in
terms of quantum coherence \cite{Zurek2}.  We introduce increased
environmental interactions into our quantum simulations in order
to test the hypothesis that the resulting behaviour will more
closely resemble the classical system.

The structure of this paper is as follows.  In
Section~\ref{sec:double-kicked rotor} we introduce the
specifically designed double-kicked rotor system, and present the
results of a classical analysis.  In Section~\ref{sec:diffusion
model} we show that this classical evolution can be well described
by a simple random model.  In Section~\ref{sec:quantum
  momentum distributions} we study the corresponding quantum
behaviour and analyze the properties of the system using the
Floquet method. In Section~\ref{sec:spontaneous} we introduce
decoherence into the quantum system by two different methods, and
also generate Wigner functions in order to help understand the
origin of classical behavior from a quantum system. Finally a
summary is contained in Section~\ref{sec:summary}. All the
parameters used in our model were chosen so as to correspond with
those used in our experiments \cite{vant,ball,christensen}.

\section{Classical Double-Kicked Rotor}\label{sec:double-kicked rotor}
The original and most commonly studied system in quantum chaos is
the $\delta$-kicked rotor.  The observation of dynamical
localization in the atomic optics realization  of the
$\delta$-kicked rotor \cite{moore95,ammann1,ammann2} provided an
important experimental link to the most studied system in
Hamiltonian chaos. However, a periodic pulsed potential of finite
time duration (as used in the $\delta$-kicked rotor experiments)
produces a KAM boundary, which becomes more noticeable for longer
pulse widths. If one wishes to study diffusion through a cantorus,
a train of single pulses is not the best system. The classical
phase space outside the first long-lived cantorus is not strongly
chaotic and contains many regular regions which will inhibit
particle diffusion.  Hence we have studied the dynamics due to a
train of double pulses. This system has been the subject of
experimental investigation by our group in its atomic optics
manifestation \cite{vant,ball,christensen}. Figure~\ref{double
pulse pic} displays our double pulse train. We can write the
dimensionless form of the Hamiltonian as
\begin{equation}
    H=\frac{p^{2}}{2}-K\cos\phi\sum_{n=-\infty}^{\infty} f(t-n)
\end{equation}
where $p$ is the dimensionless momentum conjugate to the
coordinate $\phi$ and $f(t)$ specifies the temporal shape of the
pulses. $K$ is the dimensionless `kicking strength' which is the
single parameter varied in our investigation of the classical
system. The double pulse train Hamiltonian can be written as
\begin{equation}
    H=\frac{p^{2}}{2}-K\sum_{m=-\infty}^{\infty} a_{m}\cos(\phi - 2\pi m\tau)
\end{equation}
where
$a_{m}=\frac{1}{10}$sinc$(\frac{m\pi}{20})\cos(\frac{m\pi}{10})$
(with the sinc function defined as sinc$(x)=\sin(x)/x$). Each
pulse is of width $\alpha/2$ and the leading edge separation of
the two pulses is given by $\Delta$.  The KAM boundaries at
$p=\pm10\pi$ and $\pm30\pi$ correspond to zero values for $a_{5}$
and $a_{15}$.

As discussed in \cite{vant}, below a critical kicking strength
$K=K^{\ast}$, the phase space of the classical system contains KAM
tori given approximately by the lines $p=\pm10\pi$ for our choice
of $\alpha=\Delta=0.1$.  For $K>K^{\ast}$ these break up to become
partially penetrable cantori.  Figure~\ref{fig:PoincareSection70}
shows the phase space for $K=70$.  We observe chaotic seas on
either side of the cantorus around $p=10\pi$.
Figure~\ref{fig:PoincareSection280} shows the phase space for
$K=280$. We have stronger chaos with little island structure
remaining.  The KAM torus around $p=30\pi$ is still unbroken for
this kick strength, and we find that this is the case for all
kicking strengths studied in this paper.  The strongly chaotic
seas surrounding the $p=10\pi$ cantorus provide an ideal phase
space structure for studying the transport of particles through
the barrier.

For a system kicked by finite length pulses, numerical solutions
for the motion are considerably more difficult to generate than
for $\delta$-kicks.  If the pulses are rectangular, then the
system alternates between the motion of a pendulum, and free
rotation. The solutions for pendulum motion are in terms of Jacobi
elliptic functions and elliptic integrals. In our simulations, the
elliptic integrals are efficiently evaluated with a specialized
algorithm coded in \texttt{C} \cite{numrec}, which is then
interfaced with MATLAB.

The starting point for the calculation of the momentum
distributions is an initial distribution uniform in $\phi$, and
Gaussian in $p$. For our chosen $\alpha$ and $\Delta$ we have
$\sigma_{p}=3.6\pi$ which gives $99.5\%$ of initial conditions
inside $p=\abs{10\pi}$.  We find that the results are not strongly
dependent on initial distribution, provided this proportion is
close to 1.  To obtain numerical distributions, we choose $10^5$
initial conditions randomly from this distribution, and propagate
them through 70 kick cycles.
Figure~\ref{fig:classical_waterfall_80 etc} shows examples of our
simulated distributions. Once the KAM boundaries at $p=\pm10\pi$
are broken the classical particles will eventually distribute
themselves uniformly between the $p=\pm30\pi$ tori.

\section{Diffusion Model} \label{sec:diffusion model}
A very simple model, which mimics some aspects of the classical
momentum distributions, is obtained by treating the system as
consisting of three regions where strong, homogeneous diffusion
occurs, divided by two permeable barriers. The system is enclosed
by unbroken barriers.  Applying this model to our system, we then
assume that after a short time, the distributions both inside and
outside the permeable barriers are each essentially uniform. We
also assume that over time, the populations can leak across the
penetrable barriers. The system can then be described by three
variables giving the population associated with each region.  If
we choose our initial condition so that the population in each of
the two `outer' regions is small, we expect the populations to
decay exponentially towards equilibrium values where the overall
distribution is completely uniform.

We now take the barriers to be located at $p=\pm10\pi$.  The outer
boundaries are at $\pm30\pi$, so the system is divided into three
equal parts. Each has a phase space area (in dimensionless units)
of $40\pi^{2}$. With each kick, some area of phase space $F$ is
mapped from inside the cantori to above $p=10\pi$.  This area is
referred to as the `phase space flux per kick' through the
$p=10\pi$ cantorus. Identical amounts of phase space are mapped
from above $p=10\pi$ to inside, from inside to below $p=-10\pi$,
and from below $p=-10\pi$ back inside. This follows from the
incompressibility of phase space flow in a Hamiltonian system and
the symmetry of this system in $p$.  The assumption of separate
uniform distributions inside each of the three intervals means
that the probability of a trajectory inside the cantori being
mapped outside is $2F/40\pi^{2}$.  The probability for a
trajectory outside the cantori to be mapped inside is
$F/40\pi^{2}$. Now let $P(\abs{p}<10\pi,t)$ be the probability for
a particular trajectory to be inside the cantori.  Take the
initial condition to be $P(\abs{p}<10\pi,0)=1$, so that both
outside regions are empty.  The evolution will maintain the
symmetry between them so that the probability that the trajectory
is in a particular outside region is
$P(p>10\pi,t)=P(p<10\pi,t)=(1-P(\abs{p}<10\pi,t))/2$.  The change
in the distribution caused by one kick is
\begin{eqnarray}
  P(\abs{p}<10\pi,t+1) - P(\abs{p}<10\pi,t) &=& -\frac{2F}{40\pi^{2}}P(\abs{p}<10\pi,t) +
    \frac{F}{40\pi^{2}} (1 - P(\abs{p}<10\pi,t))\\
    &=& -\frac{3F}{40\pi^{2}}(P(\abs{p}<10\pi,t)-\frac{1}{3}).
\end{eqnarray}
We can write this as
\begin{equation}
  P(\abs{p}<10\pi,t+1)-\frac{1}{3} = \left( 1-\frac{3F}{40\pi^{2}} \right)
  (P(\abs{p}<10\pi,t)-\frac{1}{3}).
\end{equation}
By induction, using our initial condition $P(\abs{p}<10\pi,0)=1$, we have
\begin{eqnarray}
    P(\abs{p}<10\pi,t)&=&\frac{1}{3}+\frac{2}{3} \left( 1-\frac{3F}{40\pi^{2}}
    \right)^{t}\\
    &=&\frac{1}{3}+\frac{2}{3} e^{a t}
\end{eqnarray}
where $a=\ln(1-3F/40\pi^{2})\approx -3F/40\pi^{2}$ ($a$ must be
small for the model to apply).  The probability for a trajectory
to be outside the cantori is given by
\begin{equation}
  P(\abs{p}>10\pi,t)=\frac{2}{3}(1 - e^{at}).
\end{equation}
The parameter $a$ is a function of kick strength which we do not
know, except that it must increase with $K$.  We can reverse the
problem, estimating the flux $F$ by fitting an exponential decay
to the proportion of trajectories outside $p=\pm10\pi$ versus $t$
for a particular kick strength.  This estimate is informative for
comparing classical and quantum results.
Figure~\ref{fig:classical_loutside}(a) shows $P(\abs{p}>10\pi,t)$
versus $t$, for a range of kicking strengths while
Figure~\ref{fig:classical_loutside}(b) shows $\ln[2/3 -
P(\abs{p}>10\pi, t)]$, which should be a straight line according
to the simple diffusion model (after a few kicks).

We see that the model appears to be reasonably good for $K\le280$.
For higher values of $K$ it may be that the transport across the
barrier is too fast for equilibrium to approximately hold in each
region.  Also, in the real system, the equilibrium population
outside will differ slightly from $2/3$, so that the straight line
form of our graph will break down for $P(\abs{p}<10\pi)$ very
close to equilibrium.  Inspection suggests that the real
equilibrium value differs from $2/3$ by $\lesssim e^{-3}\approx
0.05$. Figure~\ref{fig:classical_flux} shows estimates of the flux
per kick $F$ through the cantori based on fitting straight lines
to the curves in Figure~\ref{fig:classical_loutside}.

\section{Quantum Momentum Distributions and Floquet States} \label{sec:quantum momentum distributions}
For comparison with classical simulations, we use an initial density matrix
\begin{equation}
  \bra{m}\PDM_{0}\ket{n} = \frac{1}{A}\exp \left( -\frac{
      n^{2}\kbar^{2} }{2\sigma_{p}^{2}} \right) \delta_{m,n}
\end{equation}
where $\sigma_{p}=3.6\pi$, $A$ is a normalization constant and
$\kbar$ is Planck's constant in our dimensionless system. Again
the results are not strongly dependent on the initial spread.  We
simulate the system dynamics for up to 70 kicks.
Figure~\ref{fig:quantum_waterfall_80 etc} shows examples of these
distributions, for the same kicking strengths as were presented in
the classical case, with $\kbar=2.6$.
Figure~\ref{fig:quantum_outside}(a) shows the quantum probability
for the atom to be outside $p=\abs{10\pi}$ versus $t$, for a range
of kicking strengths. Figure~\ref{fig:quantum_outside}(b) shows
$\ln[2/3 - P(\abs{p}>10\pi, t)]$.

Comparing these graphs to Figure~\ref{fig:classical_loutside} we
see that, for all of these kicking strengths, the quantum
behaviour deviates qualitatively from the classical after less
than 20 kicks, and the probability outside the boundary for a
quantum system always levels off well below the classical
equilibrium value.  This is consistent with the premise that the
diffusion is suppressed when the phase space flux across the
cantorus is $\lesssim\kbar$.  This KAM localization is distinct
from the more widely studied dynamical localization \cite{vant}.

The time evolution operator $U$ is unitary, so that its
eigenvalues are of the form $\alpha_{j}=\exp(-iE_{j}/\kbar)$ where
$E_{j}$ is a (dimensionless) quasi-energy.  Its eigenstates
(\emph{quasi-energy states} or \emph{Floquet states}) satisfy
\begin{equation}
  U\ket{\alpha_{j}}=e^{-iE_{j}/\kbar}\ket{\alpha_{j}}.
\end{equation}
In a basis made up of these states, the evolution operator is
diagonal.  They are therefore equivalent to eigenstates of the
Hamiltonian in a time-independent system, and if we examine the
system only once per kicking cycle, they are effectively
stationary.  If we are interested in the state of the system after
quantum saturation has occurred, we can examine the asymptotic
(long-time average) momentum distribution, which can be written
\begin{eqnarray}
    P(n|\rho_{0}) &=& \lim_{N\rightarrow\infty}
    \frac{1}{N}\sum_{t=0}^{N-1}
    \bra{n}\rho_{t}\ket{n}\\
    &=& \lim_{N\rightarrow\infty}
    \frac{1}{N}\sum_{t=0}^{N-1}
    \bra{n}U^{t}\rho_{0}U^{\dagger t}\ket{n}.
\end{eqnarray}
By inserting the spectral decomposition of $U$ (representation in
terms of the Floquet states) we obtain~\cite{geisel86,geisel89}
\begin{equation}
  P(n|\rho_{0}) = \sum_{j}\bra{\alpha_{j}}\PDM_{0}\ket{\alpha_{J}}
  \cdot \abs{\expect{n|\alpha_{j}}}^{2}
\end{equation}
and if our initial condition is a pure state
$\PDM_{0}=\ket{n_{0}}\bra{n_{0}}$ we have
\begin{equation}
  P(n|n_{0}) = \sum_{j}\abs{\expect{n_{0}|\alpha_{j}}}^{2}
  \cdot \abs{\expect{n|\alpha_{j}}}^{2}.
\end{equation}
This function characterizes the asymptotic mixing between the two
momentum states.  We can generate pseudo-colour plots of
$P(n|n_{0})$, to obtain visual representations of the momentum
confinement in the quantum system for particular choices of $K$
and $\kbar$.

Figure~\ref{fig:FloquetpcolorK80 etc} shows a series of these
pseudo-colour plots.  We have used $\kbar=2.6$ and $N=128$, so
that $\kbar \abs{n_{max}}> 50 \pi$. The plots use a logarithmic
colour-scale.  Black and dark grey regions correspond to
vanishingly small probability. Mid grey regions indicate a low
probability, at the level of quantum tunneling. Pale grey and
white regions have a sizeable probability density. Inspection of
the plots shows that their shape appears to be chiefly determined
by the classical barriers to momentum diffusion.  The light line
along the line $p=p_{0}$ arises from each state being mapped onto
itself, to some degree. Each plot also shows a central light
square with $\abs{p}$ and $\abs{p_{0}}$ less than $10\pi$,
indicating the strong coupling of each state in this range to all
the others. Sharp borders on this square indicate strong
confinement by the cantori. These borders blur as the cantori
become less effective. For increased kicking strength we
eventually see a corresponding square associated with the KAM tori
at $p=\pm30\pi$. Figure~\ref{fig:FloquetpcolorK80 etc}(a) has
$K=80$ and we see very strong signatures of the classical
barriers. Figure~\ref{fig:FloquetpcolorK80 etc}(b) and (c) show
$K=180$ and $K=280$ respectively. We see significant penetration
of the $p=\pm10\pi$ cantori, but their effect is still obvious.
The $p=\pm30\pi$ KAM tori still provide a very strong barrier.
Figure~\ref{fig:FloquetpcolorK80 etc}(d) has $K=400$.  The effects
of the $p=\pm10\pi$ cantori have almost disappeared, but
penetration through the $p=\pm30\pi$ KAM tori is still fairly
small.

Again, we see strong localization which begins to break down for
$K\approx300$.  Referring to Figure~\ref{fig:classical_flux}, we
see that the classical flux through each cantorus becomes
comparable with $\kbar=2.6$ when $K\approx250$.  It has previously
been observed~\cite{brown86} that this criterion appears to give a
reasonable estimate of the kicking strength for which strong
quantum localization by a cantorus will be destroyed.

\section{Decoherence via Spontaneous Emission}\label{sec:spontaneous}

In order to study decoherence we consider the atomic optics
manifestation of this double kicked system
\cite{vant,ball,christensen}. A periodic potential is created by
two counter-propagating laser beams, and atoms are subjected to a
force due to the dipole potential. The dipole potential is derived
by neglecting the excited state amplitude of the atom
\cite{CT-85,graham92}. We will now consider the first order
effects arising from a small non-zero amplitude. Rather than
making coherent stimulated transitions which is the usual
interaction between the atoms and the pulsed potential, an atom in
the excited state may, with finite probability, decay to the
ground state by spontaneous emission. We can treat this effect as
a stochastic process, which is different for each realization. The
effect of a spontaneous emission event arises from the recoil
momentum imparted to the atom by both the photon exciting the atom
and the photon emitted by the atom.  The atom recoils with a
change in $p$ of $u\kbar$ where $-1\leq u \leq 1$ and
$u=\half(\pm1+\cos\beta)$, where the upper and lower signs occur
with equal probability and $\beta$ is the angle which the
spontaneously emitted photon makes with the $x$ axis.  $\beta$ is
random, with a distribution which is a sum of dipole distributions
over the set of possible atomic orientations.  To a fairly good
approximation \cite{ammann1}, $u$ can be treated as uniformly
distributed between $-1$ and $1$. We define $\eta$ to be the
probability per kicking cycle that the atom undergoes spontaneous
emission.

\subsection{Density Matrix Calculations}
To include spontaneous emission in our density matrix
calculations, we use the following approximate technique.  For a
particular run we choose a probability per kicking cycle $\eta$
that a particular atom will spontaneously emit.  Equivalently
$\eta$ is the proportion of the atoms in the ensemble represented
by the density matrix which will spontaneously emit in each cycle.
We then discretise $u$, so that $u=\pm1$ with equal probability.
Therefore the recoil of an atom in state $\ket{n}$ will leave it
in state $\ket{n-1}$ or $\ket{n+1}$. These states are
representable by the density matrix, unlike those arising from
continuous $u$.  Once per kick, the following replacement is made:
\begin{equation}
\bra{m}\PDM\ket{n}\leftarrow\half\eta\left(\bra{m+1}\PDM\ket{n+1}+
  \bra{m-1}\PDM\ket{n-1}\right)+(1-\eta)\bra{m}\PDM\ket{n}
\end{equation}
where we apply periodic boundary conditions.

We have also performed Monte Carlo wavefunction simulations which
are more realistic but considerably more time consuming.  In
particular they take into account the continuous distribution of
recoil momenta in the $x$ direction, and the fact that spontaneous
emission can occur at any time during the laser kick. The recoil
due to spontaneous emission is continuous between $-\kbar$ and
$\kbar$, and alters the `ladder' on which coherent dynamics occur
for the particular atom.  The approximate way in which we account
for spontaneous emission in our density matrix calculation
produces results which are negligibly different from the Monte
Carlo wavefunction calculation, but are significantly more
computationally efficient.

\subsection{Momentum Distributions}
Figures~\ref{fig:quantum SE2 waterfall} and \ref{fig:quantum SE5
waterfall} show momentum distributions versus number of kicks for
the quantum double-kicked rotor, with spontaneous emission effects
included. Figure~\ref{fig:quantum SE2 waterfall}(a) shows
behaviour for $K=80$ and $\eta=2\%$. Comparing this to
Figure~\ref{fig:quantum_waterfall_80 etc}(a), any broadening of
the distributions due to the spontaneous emission is not obvious.
However the `spiky' nature of the pure quantum version has been
significantly suppressed. Figure~\ref{fig:quantum SE5
waterfall}(a) has the same kicking strength and $\eta=5\%$. Here
we still see little broadening but even stronger suppression of
the quantum peaks. Figure~\ref{fig:quantum SE2 waterfall}(b), with
$K=180$ and $\eta=2\%$, again shows similarity to the pure quantum
case (in Figure~\ref{fig:quantum_waterfall_80 etc}(b)), except for
some suppression of peaks and slight broadening. In
Figure~\ref{fig:quantum SE5 waterfall}(b) has $K=180$ and
$\eta=5\%$. We now see unmistakable movement of probability into
the wings of the distribution, qualitatively resembling that in
the classical system for $K=180$, in
Figure~\ref{fig:classical_waterfall_80 etc}(b), although not as
strong. Figure~\ref{fig:quantum SE2 waterfall}(c) has $K=280$ and
$\eta=2\%$. There is much more flow of probability through the
cantori than in the pure quantum case in
Figure~\ref{fig:quantum_waterfall_80 etc}(b), although the
spikiness of the distribution is still strikingly different from
the classical case in Figure~\ref{fig:classical_waterfall_80
etc}(b). For $\eta=5\%$ in Figure~\ref{fig:quantum SE5
waterfall}(c), the distributions might be considered to look more
classical than quantum, although the rate of transport through the
boundary does not match the classical case. Finally for $K=400$,
we again see (Figures~\ref{fig:quantum SE2 waterfall}(d) and
\ref{fig:quantum SE5 waterfall}(d)) a significant increase in the
probability flow due to spontaneous emission, accompanied by
distribution shapes which qualitatively resemble classical
behaviour except for a somewhat smaller rate of transport.

\subsection{Wigner Functions}\label{sec:wigner functions}

A convenient way to visualize the information represented by the
density matrix is in the form of a Wigner function.  For a
discrete, truncated basis we use the toroidal Wigner function as
defined in \cite{kolovsky:96},
\begin{equation}
w(X_{k},P_{l},t)=\sum_{j=0}^{2N-1}\exp\left(i\frac{\pi
jk}{N}\right)
\frac{1+(-1)^{l+j}}{2}\bra{\frac{l+j}{2}}\PDM\ket{\frac{l-j}{2}}
\end{equation}
where $P_{l}=(\kbar/2)l$ and $X_{k}=\pi k/N$.  This gives a Wigner
function defined on a $2N\times 2N$ grid.  Averaging over cells of
four adjacent points we reduce the grid to $N \times N$. This was
implemented in \textsc{Matlab} using a fast Fourier transform
algorithm.

Figures~\ref{fig:wigner_SE0_80} through \ref{fig:wigner_SE5_400}
show a series of three-dimensional graphs of Wigner functions.
Each function represents the state of the quantum double-kicked
system after 20 kicks, with the same initial condition as for our
other simulations.  Figure~\ref{fig:wigner_SE0_80} has $K=80$ and
no spontaneous emission.  We see that the Wigner function is
strongly contained by the classical barriers.  It has a
complicated folded shape with some sharp spikes.  In
Figure~\ref{fig:wigner_SE5_80} we introduce spontaneous emission
with rate $\eta=5\%$.  The Wigner function is qualitatively
similar, but close comparison shows that it has become somewhat
smoother, with the contrast between peaks and troughs being
reduced.  In Figure~\ref{fig:wigner_SE0_180} we have $K=180$ and
$\eta=0$.  The Wigner function `spills' over the classical
barriers, although the total probability outside is small. We
chiefly notice that the Wigner function is now much more
complicated in shape, varying rapidly in position and momentum.
There are significant negative peaks present. The addition of
spontaneous emission in Figure~\ref{fig:wigner_SE5_180} again
serves to suppress this variation. It is not obvious from
inspection of these graphs, but the cantori localization is also
destroyed to some degree by finite $\eta$.

In Figure~\ref{fig:wigner_SE0_280}, with $K=280$ and $\eta=0$, we
again see an increase in the complexity of the Wigner function. At
this kicking strength, the effect of the KAM tori around
$p=\pm30\pi$ has become apparent.  The introduction of spontaneous
emission in Figure~\ref{fig:wigner_SE5_280} again serves to smooth
this variation somewhat, while pushing more (quasi) probability
into the wings of the function.  Finally we examine $K=400$.  For
the situation without spontaneous emission
(Figure~\ref{fig:wigner_SE0_400}) the function is again very
complicated and noisy-looking in shape.  The main visible change
with increased kicking strength is the increase in the function
near to the $p=\pm30\pi$ boundaries.  Introduction of spontaneous
emission (Figure~\ref{fig:wigner_SE5_400}) again suppresses the
rapid variation to some extent.

In the limit of small $\kbar$ we expect that the Wigner
pseudo-probability distribution will tend to a classical
probability distribution.  The Wigner function itself cannot be
interpreted as a probability distribution because it is not always
positive.  We can argue that the Wigner functions of states of
particularly quantum character will exhibit this non-positivity
strongly. The normalization of our discrete Wigner function is
\begin{equation}
  \sum_{k,l=1}^{N} \bar{W}(\phi_{k},p_{l}) = 1
\end{equation}
but due to non-positivity
\begin{equation}
  \sum_{k,l=1}^{N} \abs{\bar{W}(\phi_{k},p_{l})} \ge 1.
\end{equation}
We would like to quantify the `non-classicality' of a given state
with a single positive number. A possible choice is
\begin{equation}
  S = \sum_{k,l=1}^{N} \left(
    \abs{\bar{W}(\phi_{k},p_{l})}
    - \bar{W}(\phi_{k},p_{l}) \right) \ge 0.
\end{equation}
We refer to S as `quantum strangeness'.  In this paper, we will
not analyze this quantity in any detail. However, for example, a
mixed state consisting of two Gaussian wave packets centered at
$(\phi,p)=(0,\pm16\kbar)$ has $S=0.1765$, while for a
superposition state with equally weighted components of the two
wavepackets $S=0.7647$. For the initial state we use for our
simulations $S=0$. In general the larger the value of $S$, the
more non-classical the character of the state.
Figure~\ref{fig:weirdness} shows this parameter calculated for the
state of our system after 20 kicks, for several kicking strengths
$K$ and spontaneous emission rates $\eta=0$, 2\% and 5\%.

\subsection{Decoherence Versus Heating}\label{sec:heating}
An atom which undergoes spontaneous emission receives a random
momentum kick. This will obviously lead to an increase in the
width of the momentum distribution,  the net effect being some
incoherent absorption of energy from the laser beams.  This effect
is referred to as `heating'.  This however is not the main
physical mechanism behind the increased diffusion we observe due
to spontaneous emission. In Figure~\ref{fig:add_outside2 5}(a) and
\ref{fig:add_outside2 5}(b) we show the additional diffusion
through $p=\pm10\pi$ over the pure quantum case, for several
kicking strengths and two values of spontaneous emission rate
$\eta$. We note that the contribution of heating to the crossing
of this boundary will be strongest when the momentum distribution
is steep near $p=\pm10\pi$.  If the broadening effect was due
entirely to heating it should be most pronounced for the smaller
kicking strengths $K$ where the pure quantum distribution is
strongly localized by the cantori, and for fixed $\eta$, should
not increase with increasing $K$. Figure~\ref{fig:add_outside2
5}(a) shows that the additional diffusion caused by spontaneous
emission rate $\eta=2\%$ is much larger at $K=180$, 280 and 400,
than at $K=80$. Figure~\ref{fig:add_outside2 5}(b) shows the same
trend for $\eta=5\%$. There must therefore be another mechanism
which increases the diffusion rate, and which is much stronger
than the heating mechanism for $K\ge180$.  This mechanism is the
destruction of quantum coherence, or decoherence, caused by the
randomizing effect of spontaneous emission.  We have also
experimentally verified that heating is negligible for
$K>150$~\cite{kendrathesis}.

\subsection[Measurement Decoherence, or the Anti-Zeno Effect]{Measurement
  Decoherence,\\ or the Anti-Zeno Effect}

Our model for spontaneous emission leads to decoherence which we
can consider to be `environmentally induced'.  Kaulakys and
Contis~\cite{kaulakys97} discuss the effect of projective momentum
measurements on the dynamics of the quantum $\delta$-kicked rotor.
They find that if a momentum measurement is made after every kick,
then quantum localization is completely destroyed, and unbounded
diffusion occurs, with the same diffusion constant as in the
classical system. They refer to this modification of the dynamics
as an \emph{anti-Zeno effect}. Each measurement corresponds to the
diagonalization of the density matrix in the momentum
representation (i.e.\ off-diagonal elements become zero), or
equivalently, the loss of all information about position. This
effect is formulated without appealing to a collapse of the
wave-function, and can in principle be produced experimentally.

We perform a similar simulation for our system to compare this
form of decoherence to the spontaneous emission induced
decoherence and to the classical motion.  Computationally this is
very similar to the density-matrix spontaneous emission
simulations.  After each double-kick coherent cycle, the density
matrix is replaced by a matrix with the same diagonal entries, but
all zero off-diagonal entries.  The cycle then repeats. We note
that the Wigner functions for states generated by this type of
simulation are very simple, being the product of a uniform
position distribution and the momentum distribution.  We therefore
do not show them, but point out that the `quantum strangeness'
parameter $S$ is always zero for these functions.

\subsection{Comparison of Quantum and Classical Results} \label{sec:comparison}
To compare our quantum and classical results, we calculate the
probability for a particle to cross the $p=\pm10\pi$ cantori,
$P(\abs{p}> 10\pi,t)$, for classical, pure quantum, spontaneous
emission and anti-Zeno simulations of the system.
Figure~\ref{fig:everything_80}(a) has $K=80$.  Classical and pure
quantum simulations give essentially horizontal lines with some
fluctuation, because the KAM tori present at this kicking strength
are effective barriers for both systems.  The initial conditions
determine the level of this line. The quantum simulations with
spontaneous emission show a gradual leakage and the quantum
anti-Zeno simulation gives a qualitatively similar result. This
suggests that, at this kicking strength, decoherence breaks down
the quantum cantorus localization, \emph{destroying} the
correspondence between the quantum and classical systems.

In Figure~\ref{fig:everything_80}(b), we present simulations for
$K=180$. There is now a clear difference between the classical and
quantum simulations.  In fact they appear to differ even for very
small $t$, where the quantum curve initially rises more sharply
than the classical, before saturation sets in.  Adding spontaneous
emission destroys the saturation, and for $t>10$ causes the
quantum system to qualitatively more closely resemble the
classical. The anti-Zeno simulation however suggests that the
limit of large decoherence is again \emph{faster} transport across
the cantori than in the classical case.  We note that this
simulation does not involve heating effects, so the transport must
be due to decoherence.

Now shifting our attention to $K=280$, in
Figure~\ref{fig:everything_80}(c), we see a large difference
between classical and pure quantum behaviour which appears to be
bridged by the introduction of spontaneous emission.  The
anti-Zeno calculation now closely corresponds to the classical
picture, and we conclude that at this kicking strength decoherence
does make the quantum system `more classical'.

Finally we consider $K=400$ in Figure~\ref{fig:everything_80}(d).
The pure quantum case is not as far from the classical behaviour
as for $K=280$, but again decoherence is effective in increasing
the similarity between the systems.  The anti-Zeno calculation now
gives results that are almost indistinguishable from the
classical.  The close correspondence between the anti-Zeno and
classical simulations for this and the previous figure can be
related to the fact that for these kicking strengths the size of
the classical flux per kick (in Figure~\ref{fig:classical_flux})
has become comparable to our dimensionless Planck's constant
$\kbar=2.6$.

Our group has previously published experimental results for
$K=280$ \cite{ball}.  These show good agreement with the
spontaneous emission simulations, especially the Monte Carlo runs.
With the elimination of some systematic experimental errors we
expect that still better agreement would be achieved.

Figure~\ref{fig:everything_80} demonstrates that, as we have seen,
when $\eta=0$, quantum saturation sets in by $t=20$ over our
entire range of kicking strengths. If we examine the $\eta=0$
results we see that the quantum strangeness $S$ is very small for
$K=80$, but rapidly increases with increasing kicking strength. In
general, we expect to see $S$ increase with the chaoticity of the
corresponding classical system.  Classical systems with strong
chaos rapidly develop phase space structure on all scales, which
cannot be reproduced in a quantum system with finite $\kbar$, so
the quantum system must begin to exhibit non-classical features.
It is interesting here that when $K=80$, $S$ is almost negligible,
which seems to correspond to the fact that the classical and
quantum systems are both strongly localized by the KAM tori in the
classical system. As $K$ increases, the small scale phase space
structure making up the cantori must differ in the classical and
quantum systems, and we see that $S$ rises quickly with $K$ in
this regime.  Examining the effects of introducing spontaneous
emission, we see that $S$ is significantly reduced even by a rate
$\eta=2\%$.  Along with the `more classical' diffusion seen in
Figure~\ref{fig:everything_80}, we can see that the Wigner
function also exhibits more `classicality' when we introduce
environmental decoherence.

\section{Discussion} \label{sec:summary}
We have analyzed and numerically simulated the classical
double-kicked rotor system and verified that it possesses KAM tori
and cantori which present barriers to diffusion.  We have
established the success of a simple diffusion model for the system
and used it to estimate the phase space flux per kick through a
cantorus as a function of kicking strength. The double-kicked
rotor is an interesting chaotic system, which would reward further
analysis. One aspect would be locating periodic trajectories of
the system, especially the series of these trajectories converging
to the noble KAM torus or cantorus.

The quantum double-kicked system shows strong localization
corresponding to classical KAM tori.  The system is also localized
by classical \emph{cantori} and does not show the sharp transition
shown classically.  Whereas the classical system eventually
reaches a uniform distribution in phase space, the quantum system
saturates with probability still significantly confined by the
classical barriers. This saturation is confirmed by a Floquet
analysis of asymptotic momentum distributions.  The effect is
still obvious even when the size of the classical flux per kick
becomes of the order of our dimensionless Planck's constant,
$\kbar$.  In addition to the saturation, we observe fluctuating
peaks in the momentum distributions, which contrast strongly to
the flat-topped distributions seen classically.  The quantum
transport through the classical boundary more closely resembles
the classical situation as the kicking strength increases, but
examination of the Wigner functions shows that the system becomes
\emph{more} non-classical. This is consistent with the general
theory of quantum chaos which suggests that stronger chaos in the
classical system accelerates the appearance of quantum coherence
effects~\cite{ammann1,kolovsky:96,Zurek1}.

The modern theory of environmental decoherence states that the
interaction of quantum systems with the environment is a necessary
condition for the appearance of classical behaviour in real
systems.  The traditional semi-classical limit, which in our
formulation is given by $\kbar\rightarrow0$, is unsatisfactory
because the quantum break time becomes infinite only
logarithmically, and can be surprisingly short for a real
macroscopic system.  After this time arbitrary quantum
superpositions may arise.

We have observed the effects of decoherence on our quantum system,
using two models; dissipation by spontaneous emission caused by
the kicking laser beam, and the anti-Zeno effect.  Where the
classical flux per kick through the cantorus is finite,
decoherence disrupts quantum saturation and the system is
qualitatively more like the classical version.  Examination of the
Wigner function also indicates that quantum interference effects
are suppressed by the decoherence. The anti-Zeno calculations are
a kind of extreme decoherence limit, and we have seen that they
reproduce the classical dynamics accurately, provided that $\kbar$
is not too large compared to the classical flux per kick.  It
appears not unreasonable to suppose that this correspondence
continues indefinitely.

There are numerous aspects of decoherence in this system which
could be further investigated.  The introduction of noise into the
kicking of the system is expected to have a similar effect to that
of dissipation through spontaneous emission~\cite{Raizen4}.
Particular states of the system will be `resistant' to
decoherence.  These are expected to be the `classical' states in
the limit of strong decoherence and small $\kbar$.  We would like
to determine some of these states and quantify the `decoherence
times' for other states, for comparison with the quantum break
time of the coherent system.  It would be interesting to
quantitatively compare the anti-Zeno effect with decoherence via
noise and dissipation.

Our results support the idea that the apparently classical nature
of the universe arises entirely from quantum mechanics.  In the
real world the unpredictable behaviour of chaotic systems must
ultimately arise from quantum uncertainty.  As further progress is
made in the investigation of decoherence, there is hope that
physics will develop a consistent picture of a smooth transition
between quantum and classical descriptions of reality.

\section*{Acknowledgements}
This work was supported by the University of Auckland Research
Committee and the Royal Society of New Zealand Marsden Fund.


\begin{figure}[p]
\begin{minipage}{\textwidth}
   \caption{Double pulse showing definitions of $\alpha$ and
   $\Delta$.}
    \label{double pulse pic}
\end{minipage}
\begin{minipage}{\textwidth}
    \caption{Poincare section for double-kicked rotor with $K=70$.}
    \label{fig:PoincareSection70}
\end{minipage}
\begin{minipage}{\textwidth}
    \caption{Poincare section for double-kicked rotor with $K=280$.
    The cantori at $p=\pm10\pi$ are no longer visible in the phase space at this resolution.}
  \label{fig:PoincareSection280}
\end{minipage}
\begin{minipage}{\textwidth}
    \caption{Classical momentum distribution as a function of
    number of kicks $t$, for the standard double-kicked system
    with (a) $K=80$, (b)$180$, (c)$280$ and (d)$400$.}
    \label{fig:classical_waterfall_80 etc}
\end{minipage}
\begin{minipage}{\textwidth}
   \caption{(a) Proportion of trajectories outside $p=\pm10\pi$,
    versus number of kicks $t$.  Kicking strength for each
    curve is indicated by the legend. (b) Logarithmic plot of
    $2/3-P(\abs{p}>10\pi)$, versus number of kicks $t$.}
    \label{fig:classical_loutside}
\end{minipage}
\begin{minipage}{\textwidth}
   \caption{Estimate of phase space flux per kick (F) through cantorus versus kicking
      strength~$K$.}
    \label{fig:classical_flux}
\end{minipage}
\begin{minipage}{\textwidth}
    \caption{Quantum momentum distribution as a function of number of kicks
     $t$, with (a) $K=80$, (b)$180$, (c)$280$ and (d)$400$.}
    \label{fig:quantum_waterfall_80 etc}
\end{minipage}
\begin{minipage}{\textwidth}
   \caption{(a) Probability outside
      $p=\pm10\pi$, versus number of kicks $t$. Kicking strength $K$ for
      each curve is $120$ (dotted), $150$ (dashed), $180$ (dot-dashed), $210$ (solid),
      $250$ (heavy dotted) and $280$ (heavy dashed).  (b) Logarithmic plot of
      $2/3-P(\abs{p}>10\pi)$, versus number of kicks $t$.
      Classically these curves are almost straight lines.}
    \label{fig:quantum_outside}
\end{minipage}
\begin{minipage}{\textwidth}
    \caption{Pseudo-colour plot of asymptotic momentum
        distribution $P(p|p_{0})$ for (a) $K=80$, (b)$180$, (c)$280$ and (d) $400$, $\kbar=2.6$. Computed
        using 128 states, upon which $P(p|p_{0})$ is normalized to 1.}
      \label{fig:FloquetpcolorK80 etc}
\end{minipage}
\begin{minipage}{\textwidth}
    \caption{Quantum momentum distribution as a function of number
    of kicks $t$ with $\eta=2\%$ (a) $K=80$, (b)$180$, (c)$280$ and (d)$400$.}
    \label{fig:quantum SE2 waterfall}
\end{minipage}
\begin{minipage}{\textwidth}
    \caption{Quantum momentum distribution as a function of number
    of kicks $t$ with $\eta=5\%$ (a) $K=80$, (b)$180$, (c)$280$ and (d)$400$.}
    \label{fig:quantum SE5 waterfall}
\end{minipage}
\begin{minipage}{\textwidth}
    \caption{Wigner function after 20 kicks with $K=80$ and no spontaneous emission.}
    \label{fig:wigner_SE0_80}
\end{minipage}
\begin{minipage}{\textwidth}
    \caption{Wigner function after 20 kicks with $K=80$ and $\eta=5\%$.}
    \label{fig:wigner_SE5_80}
\end{minipage}
\begin{minipage}{\textwidth}
    \caption{Wigner function after 20 kicks with $K=180$ and $\eta=0$.}
    \label{fig:wigner_SE0_180}
\end{minipage}
\begin{minipage}{\textwidth}
        \caption{Wigner function after 20 kicks with $K=180$ and $\eta=5\%$.}
     \label{fig:wigner_SE5_180}
\end{minipage}
\end{figure}

\begin{figure}[p]
\begin{minipage}{\textwidth}
    \caption{Wigner function after 20 kicks with $K=280$ and $\eta=0$.}
    \label{fig:wigner_SE0_280}
\end{minipage}
\begin{minipage}{\textwidth}
    \caption{Wigner function after 20 kicks with $K=280$ and $\eta=5\%$.}
    \label{fig:wigner_SE5_280}
\end{minipage}
\begin{minipage}{\textwidth}
    \caption{Wigner function after 20 kicks with $K=400$ and $\eta=0$.}
    \label{fig:wigner_SE0_400}
\end{minipage}
\begin{minipage}{\textwidth}
    \caption{Wigner function after 20 kicks with $K=400$ and $\eta=5\%$.}
    \label{fig:wigner_SE5_400}
\end{minipage}
\begin{minipage}{\textwidth}
    \caption{`Quantum strangeness' $S=\sum_{k,l=1}^{N} \left(\abs{\bar{W}(\phi_{k},p_{l})} -
        \bar{W}(\phi_{k},p_{l}) \right)$ versus kicking strength $K$
      for spontaneous emission rates $\eta=0$, 2\%, 5\% as indicated
      in legend.}
    \label{fig:weirdness}
\end{minipage}
\begin{minipage}{\textwidth}
    \caption{(a) Relative effect of spontaneous emission rate $\eta=2\%$ for
      different kicking strengths $K=80$ (solid), $180$ (dotted), $280$
      (dashed) and $400$ (dot-dashed).
      Proportion of atoms outside $p=\pm10\pi$ for quantum simulations
      with $\eta=0$ and $\eta=2\%$ spontaneous emission are calculated
      and the difference is plotted. (b) same as (a) but with $\eta=5\%$.}
    \label{fig:add_outside2 5}
\end{minipage}
\begin{minipage}{\textwidth}
    \caption{Comparison of classical and quantum simulations with
      (a) $K=80$, (b) $180$, (c) $280$ and (d) $400$.
      Probability of finding a given atom outside $p=\pm10\pi$ is
      plotted versus number of kicks $t$.  Coherent quantum
      evolution (light solid line);  Quantum evolution with
      spontaneous emission rates $\eta=2\%$ (dashed line) and $5\%$
      (dash-dotted line) respectively.  Quantum simulation with
      anti-Zeno effect (dotted line).  Classical simulation (heavy solid line).}
    \label{fig:everything_80}
\end{minipage}
\end{figure}

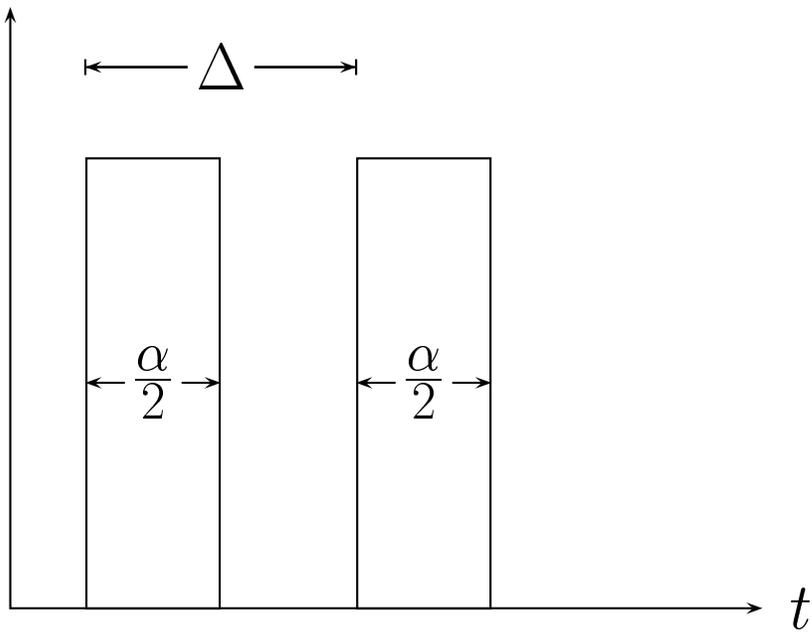
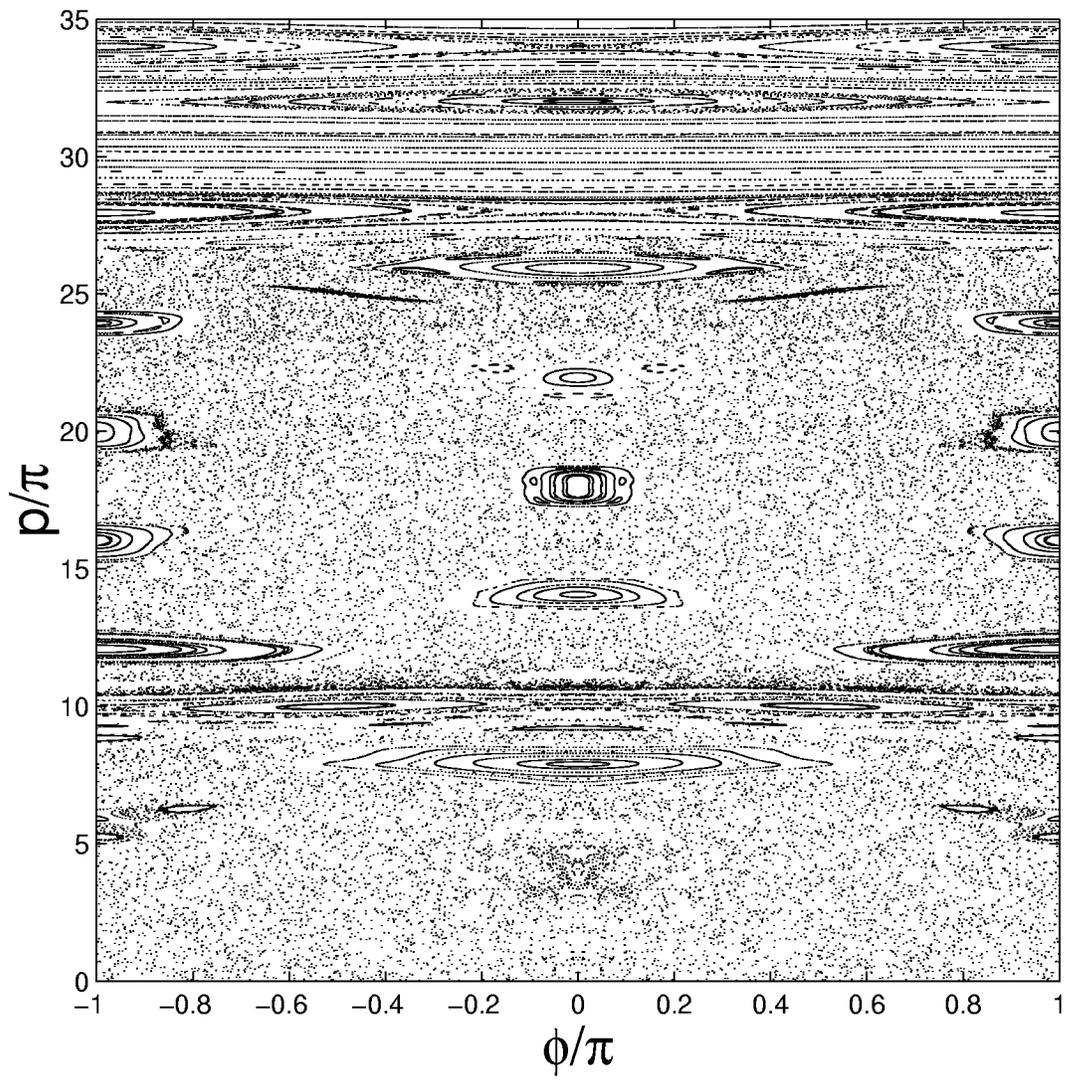
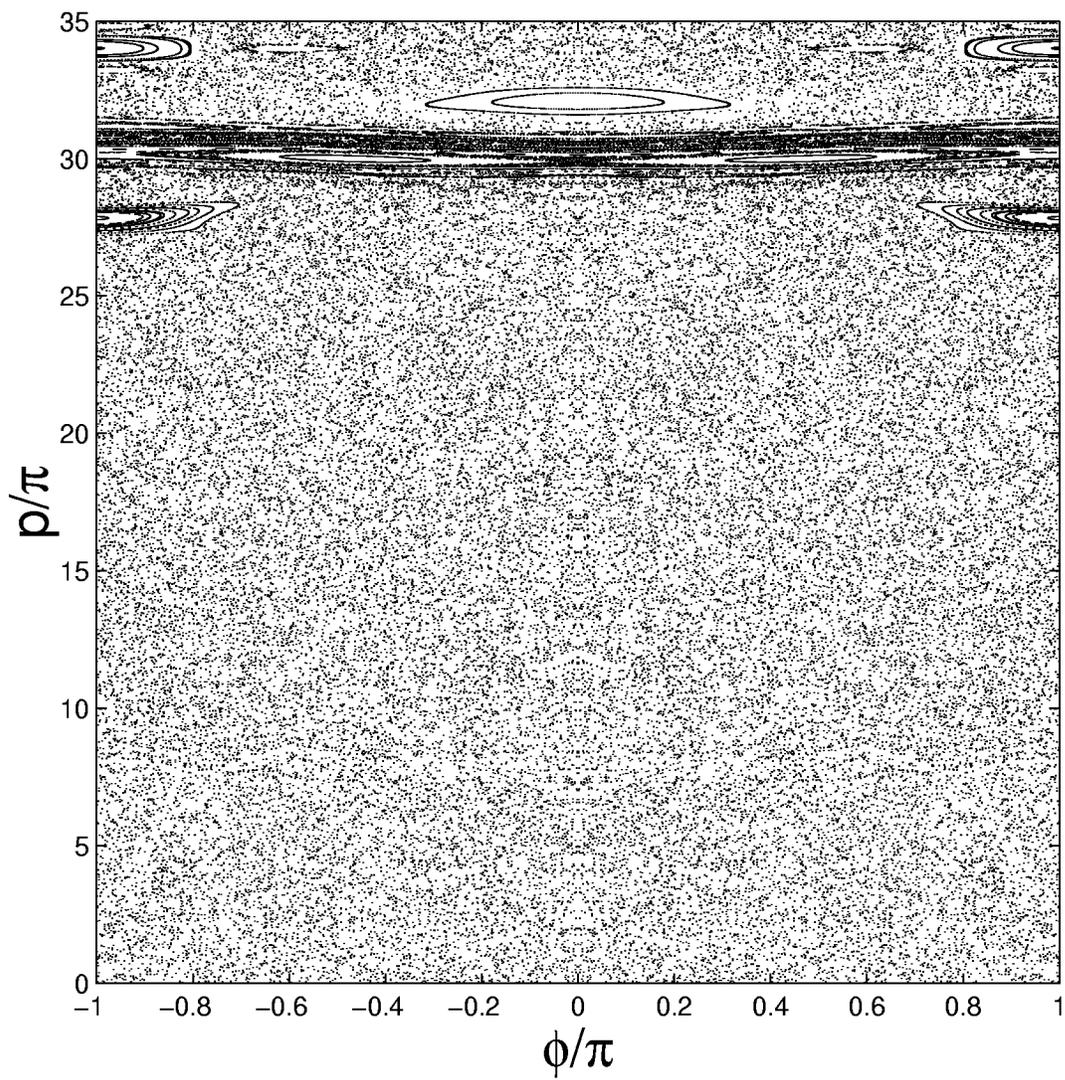
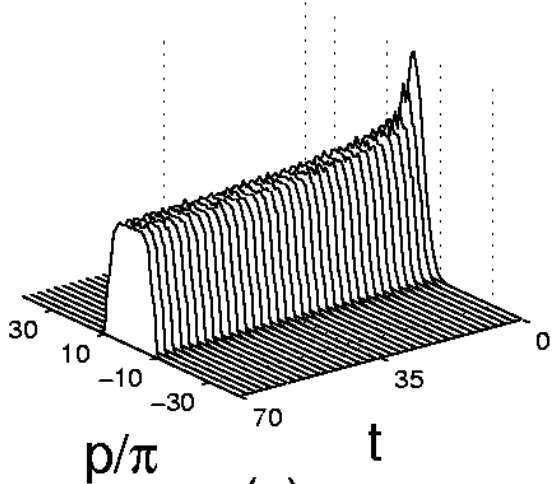
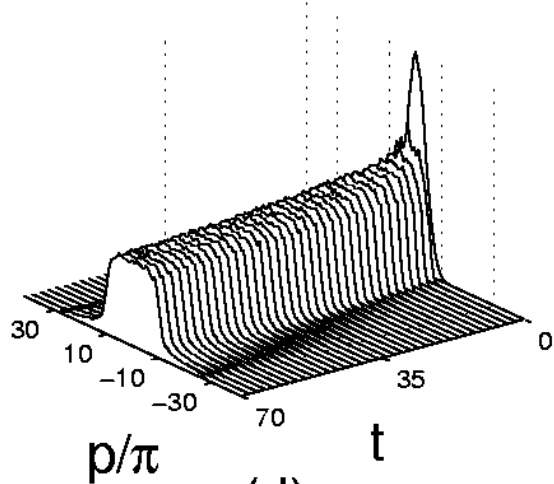
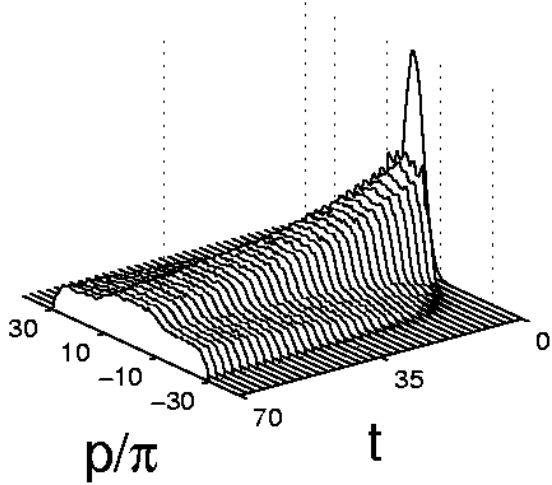
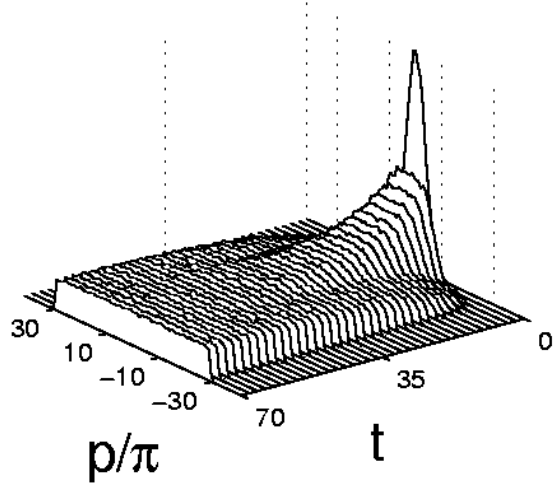
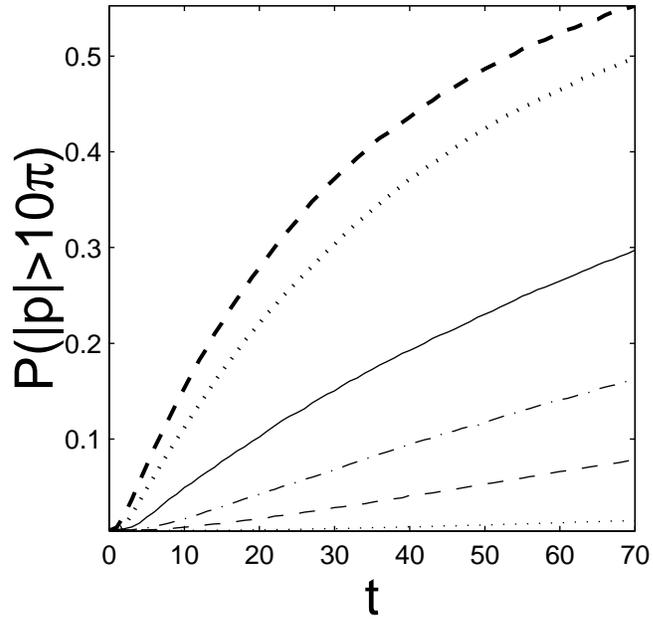
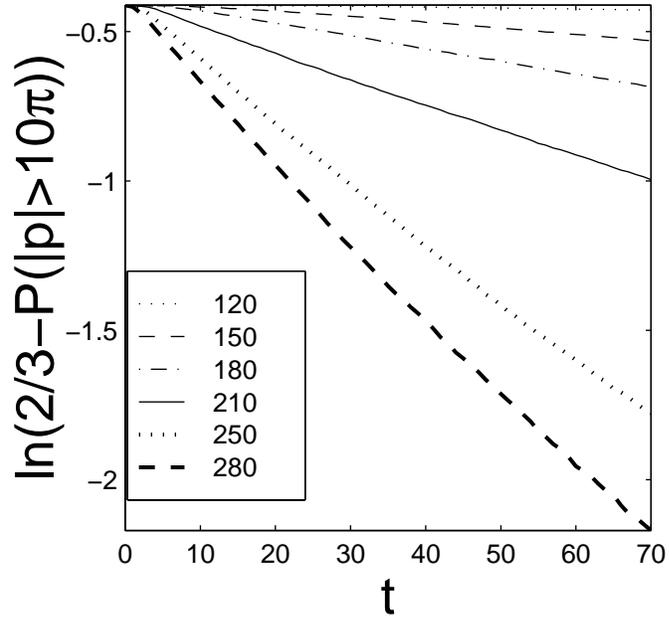
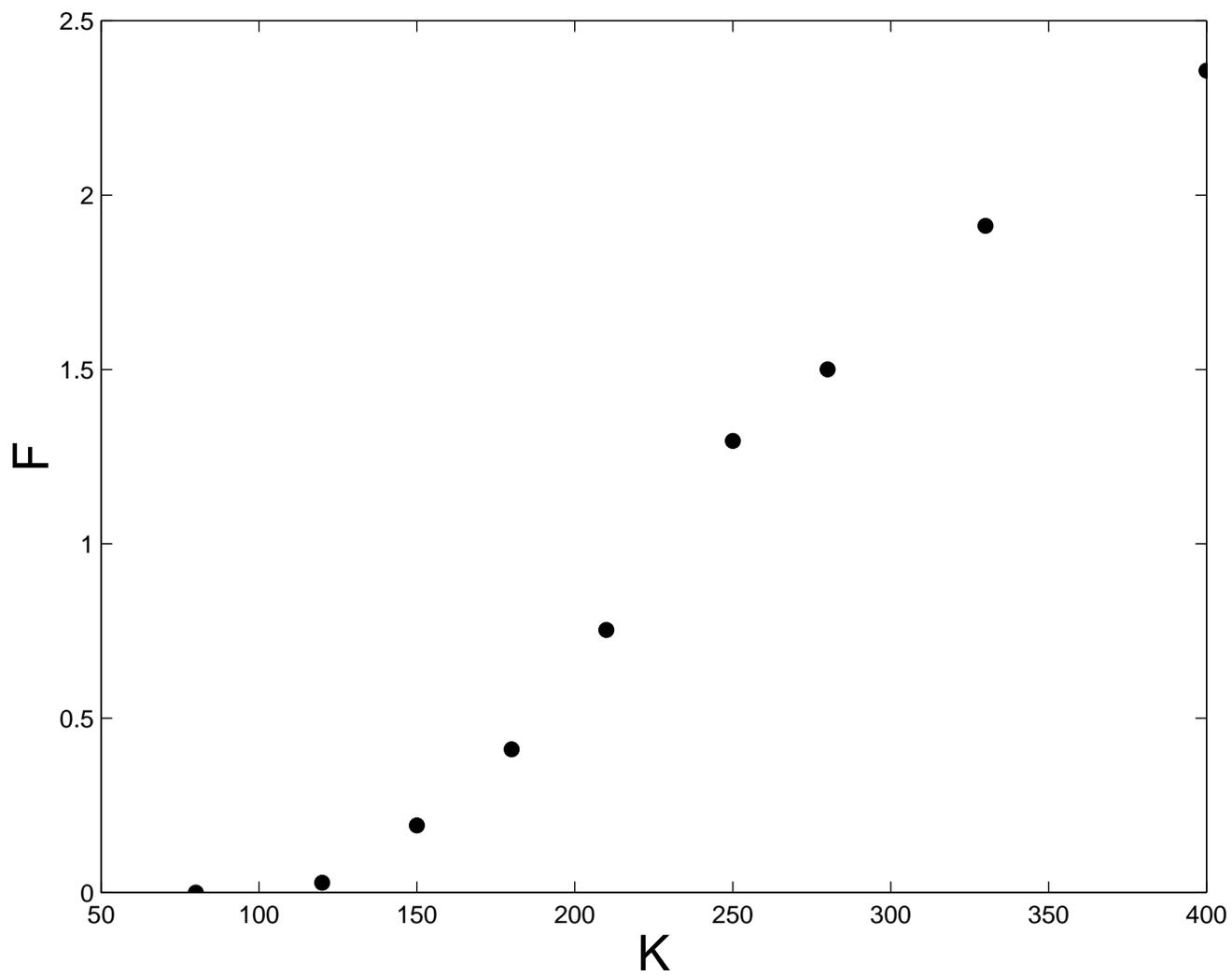
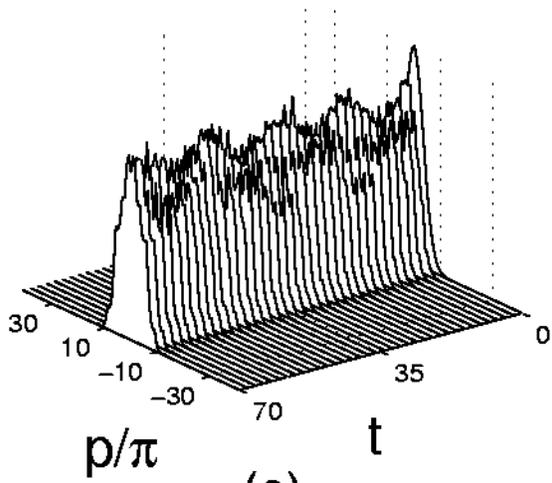
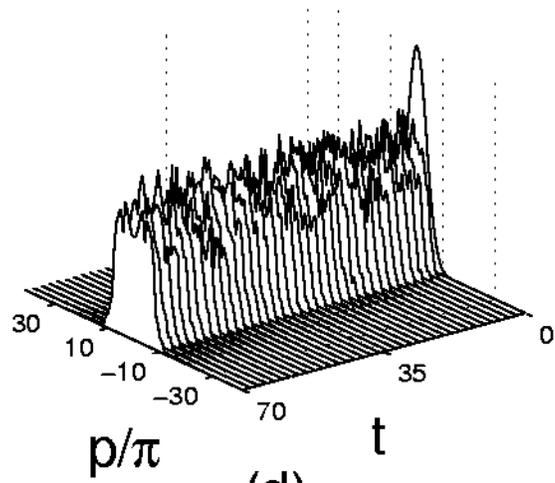
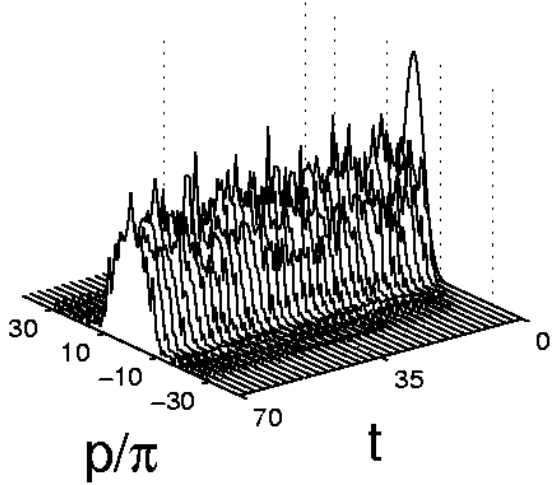
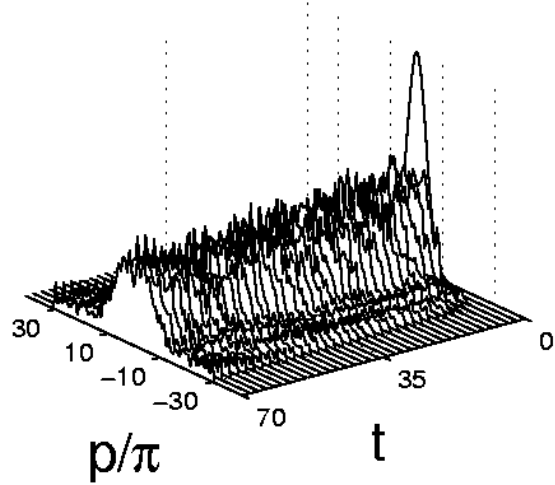
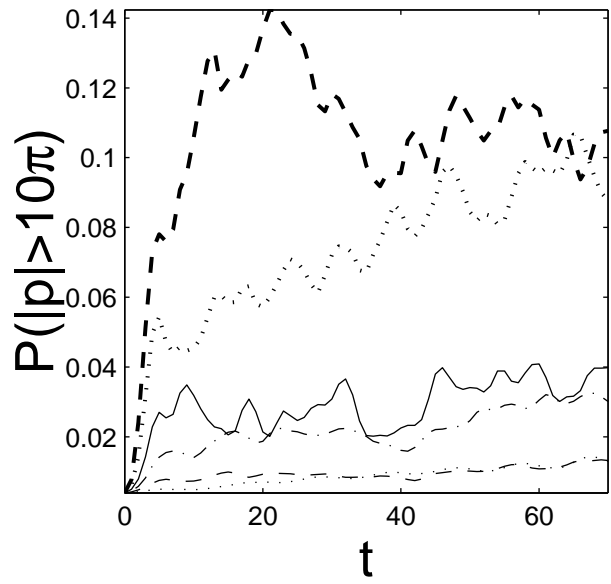
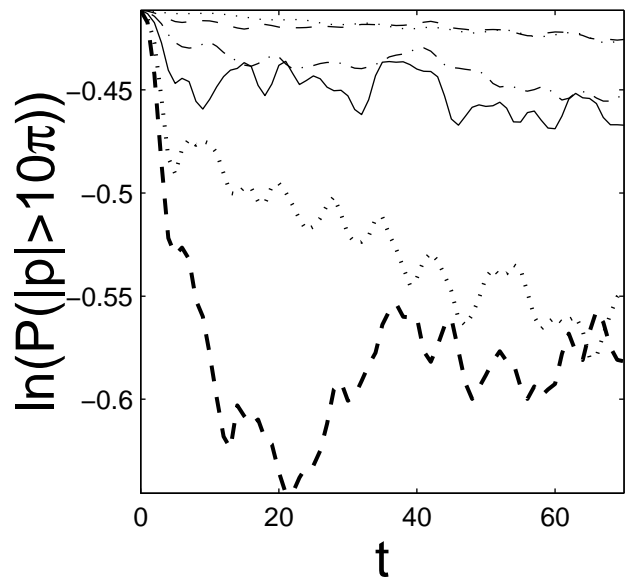
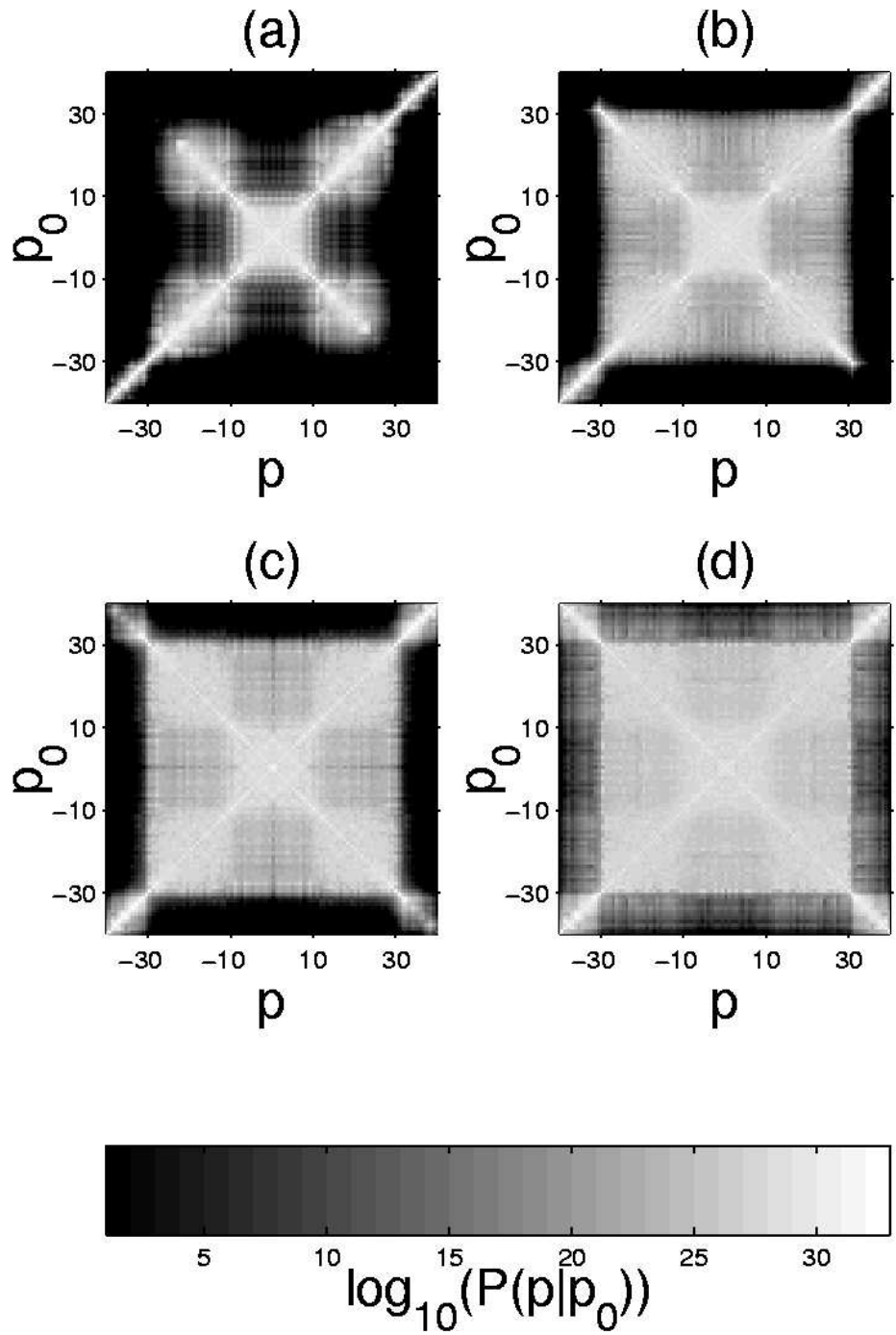
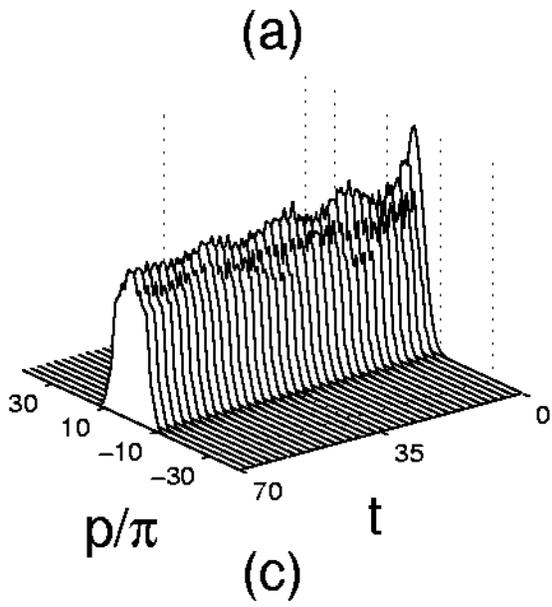
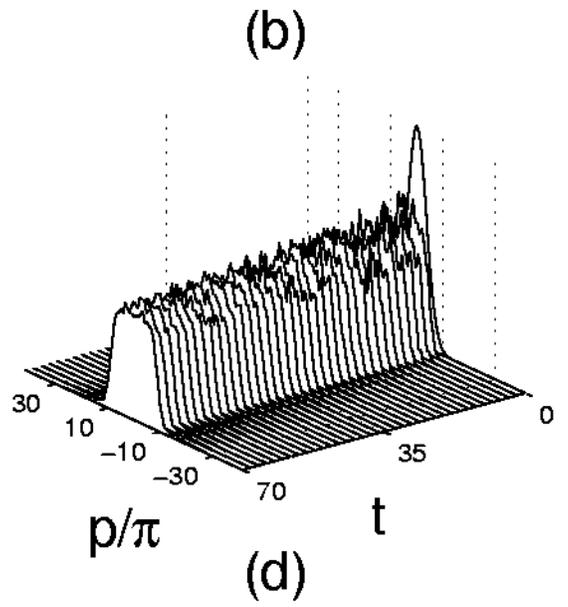
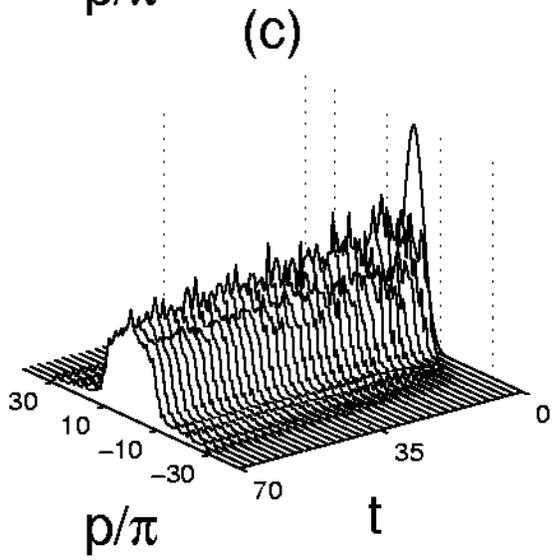
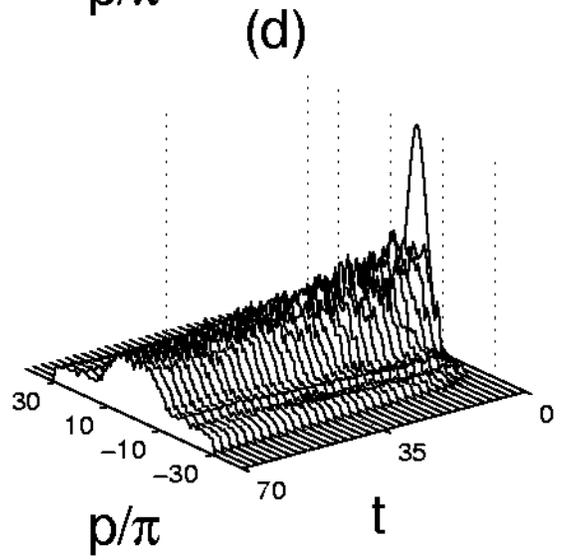
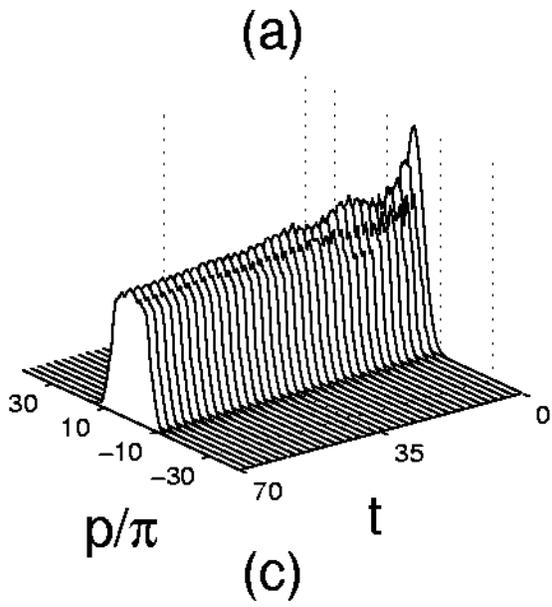
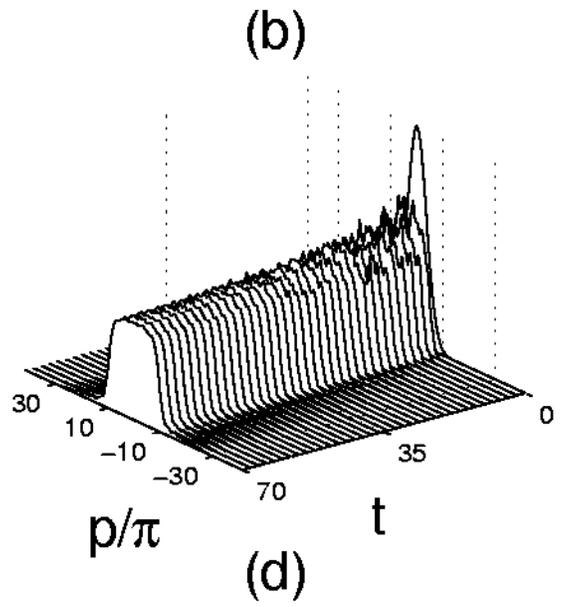
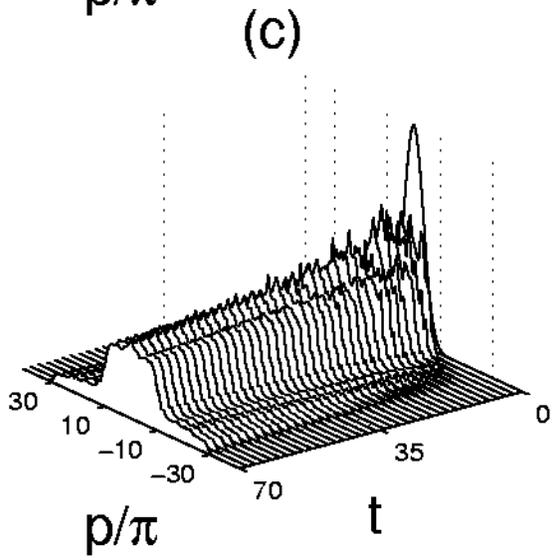
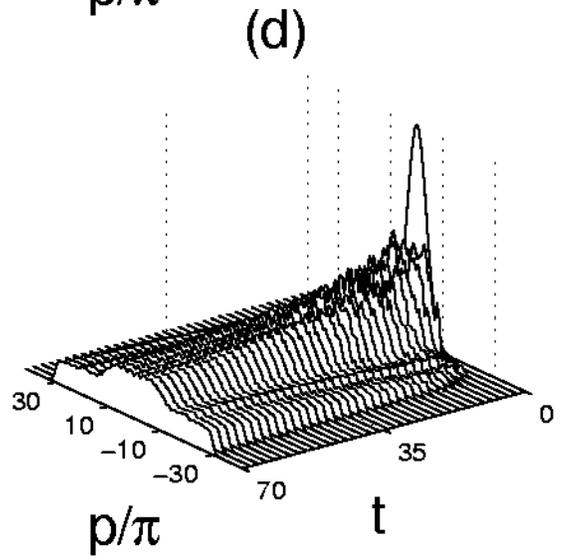
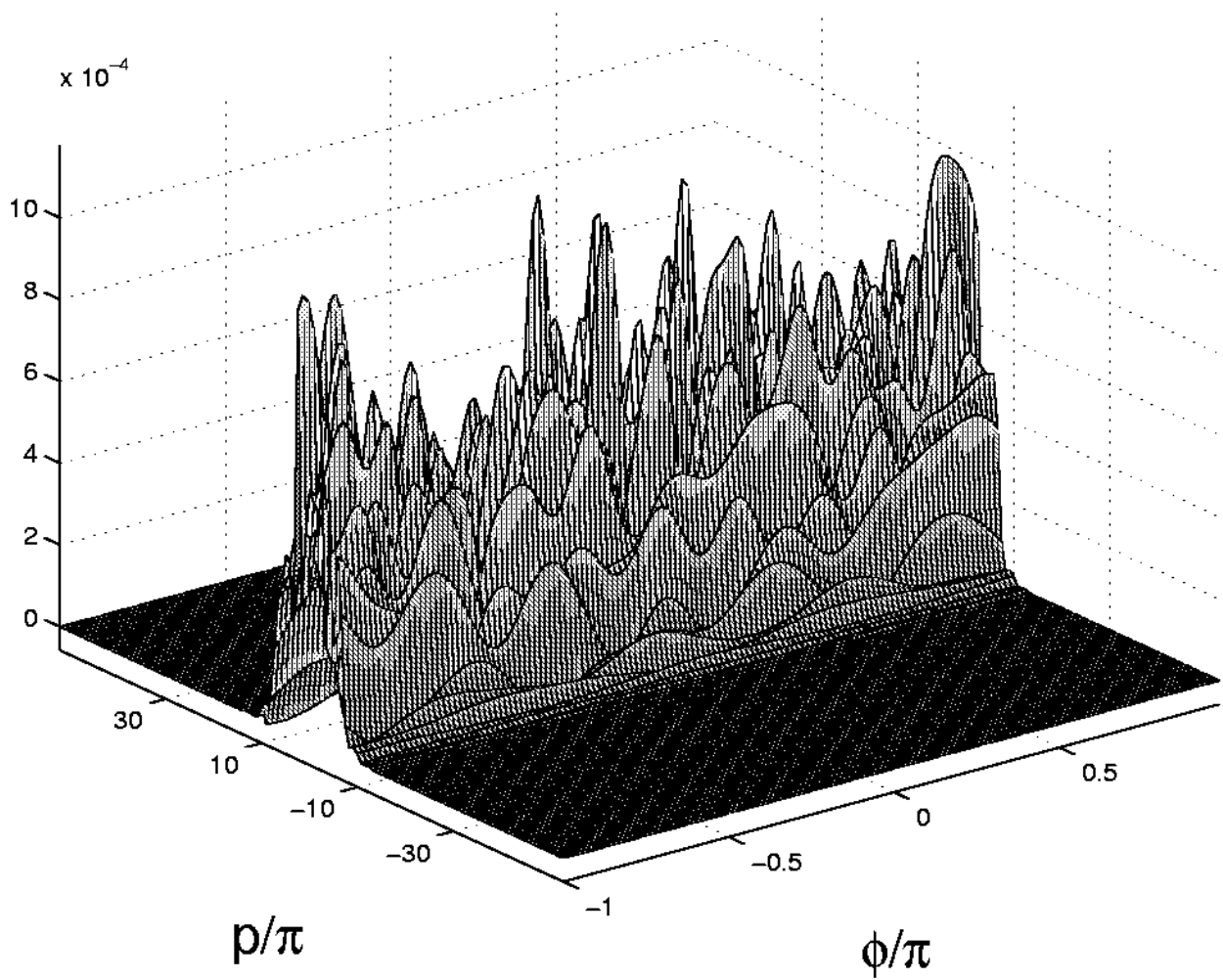
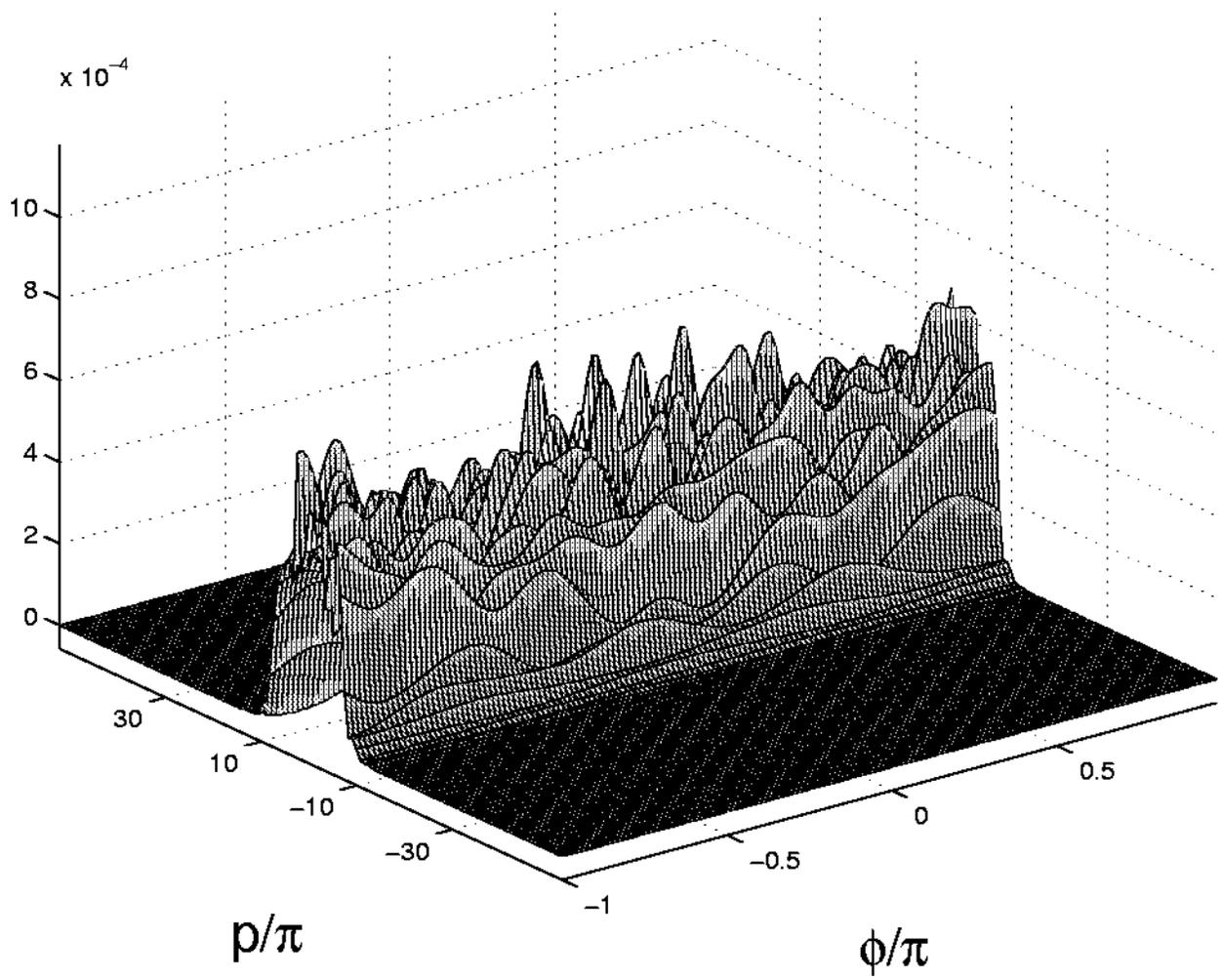
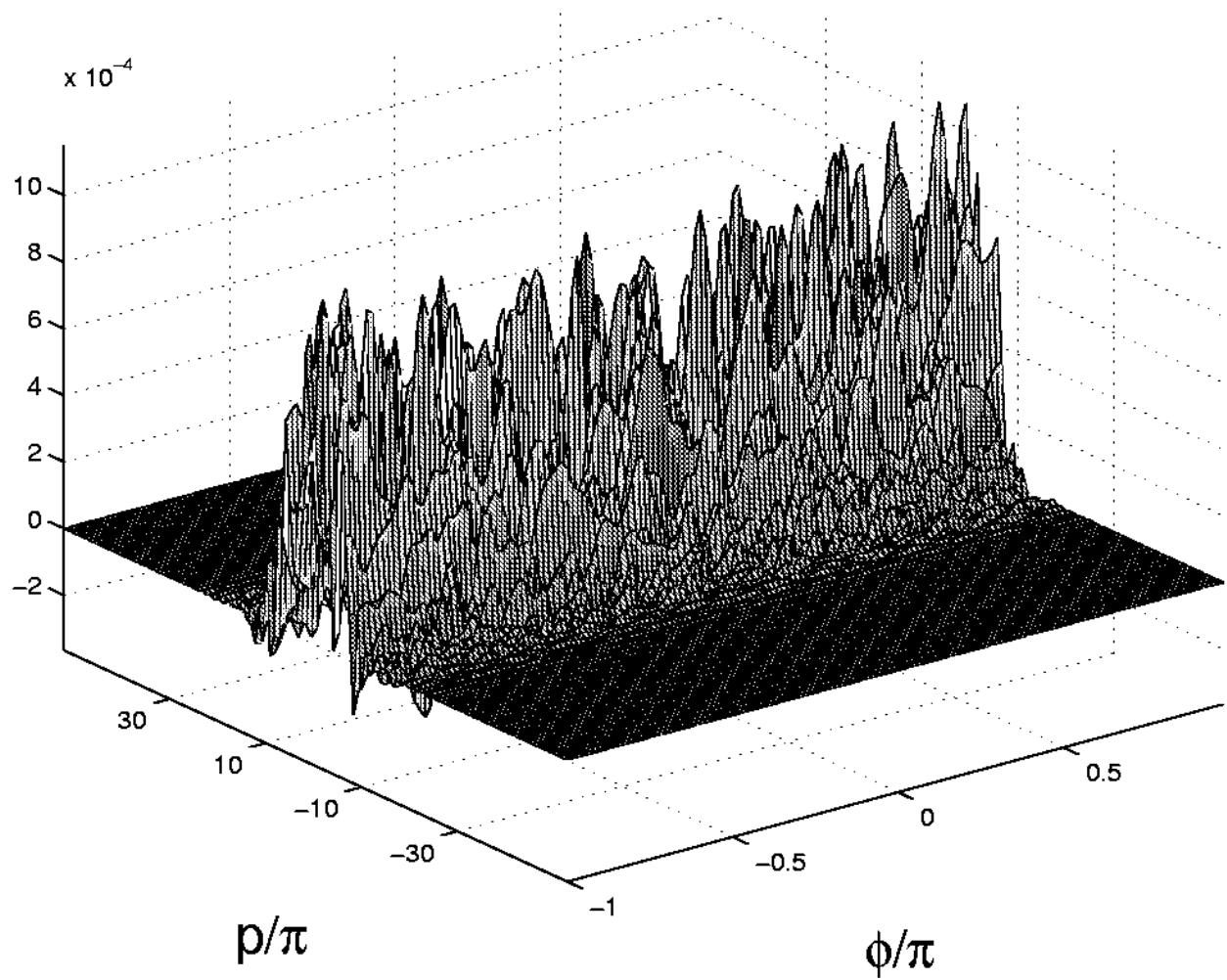
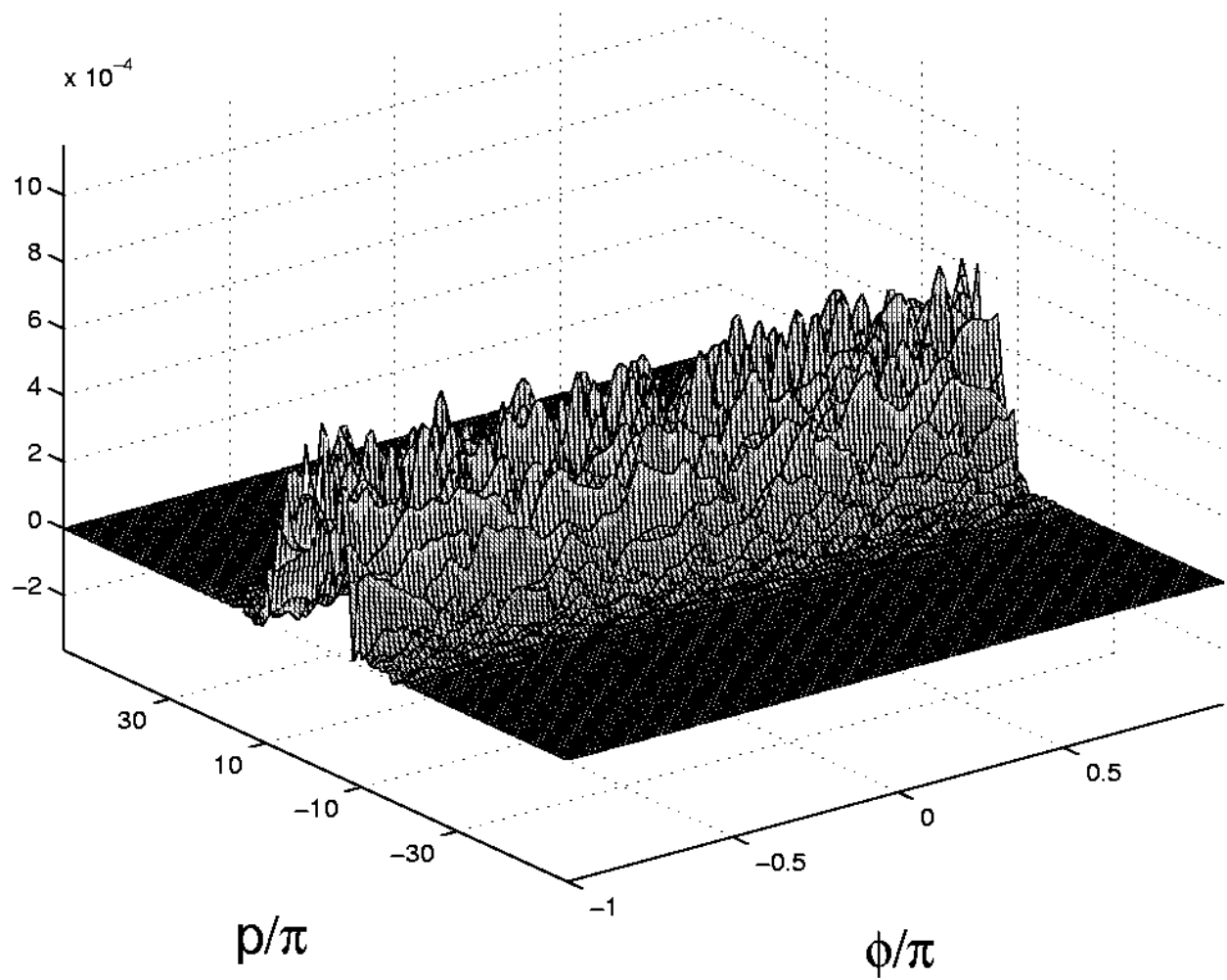
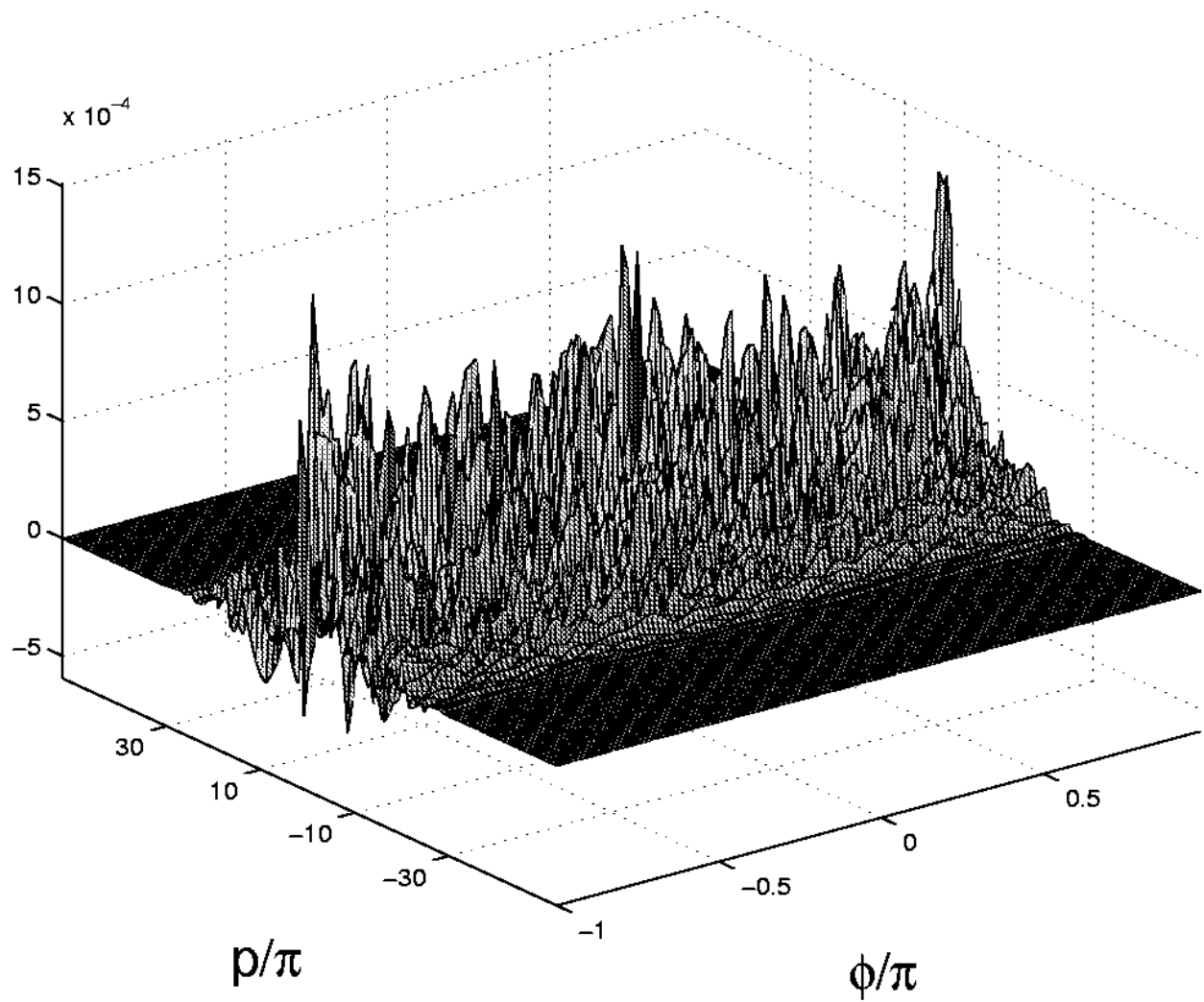
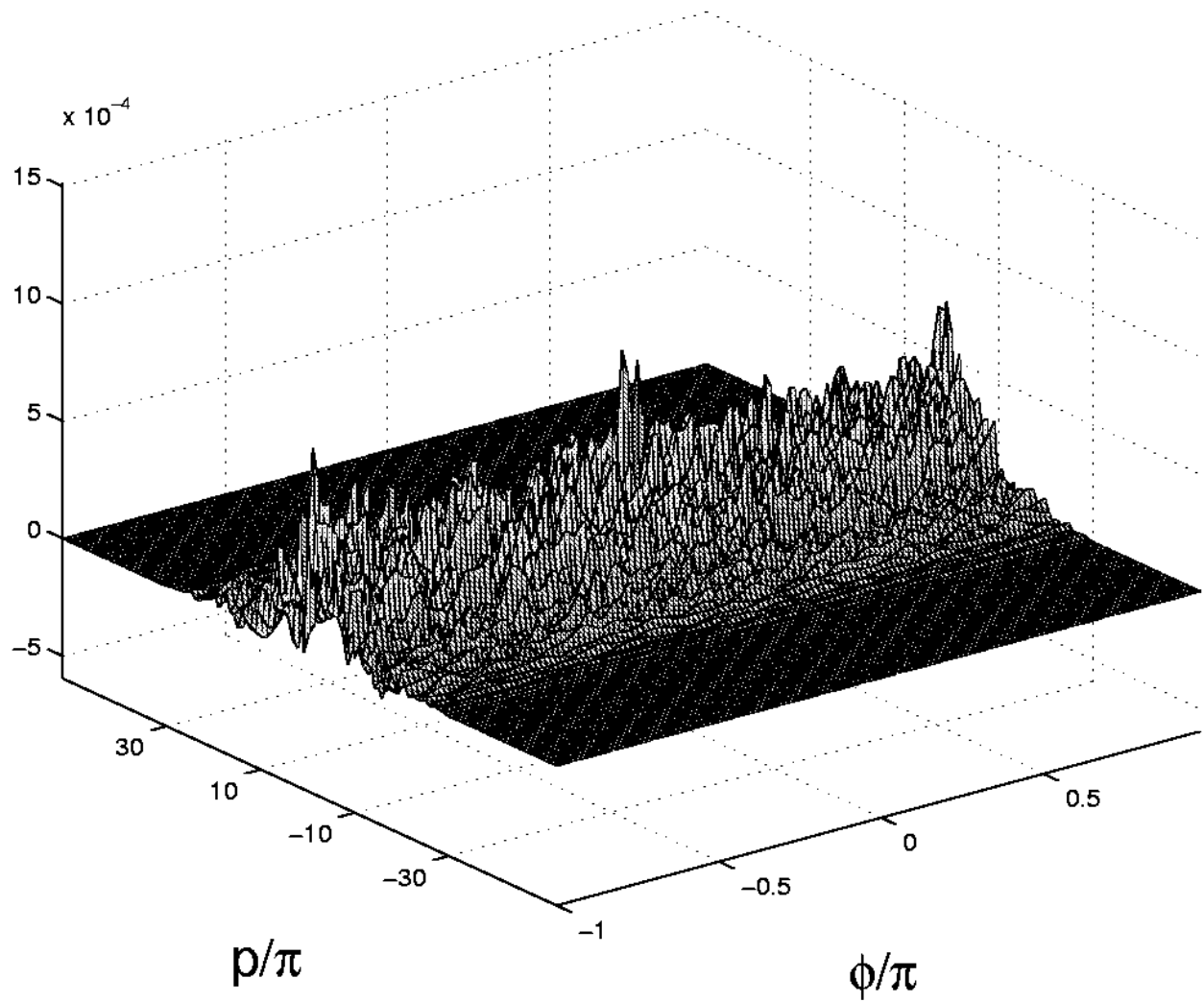
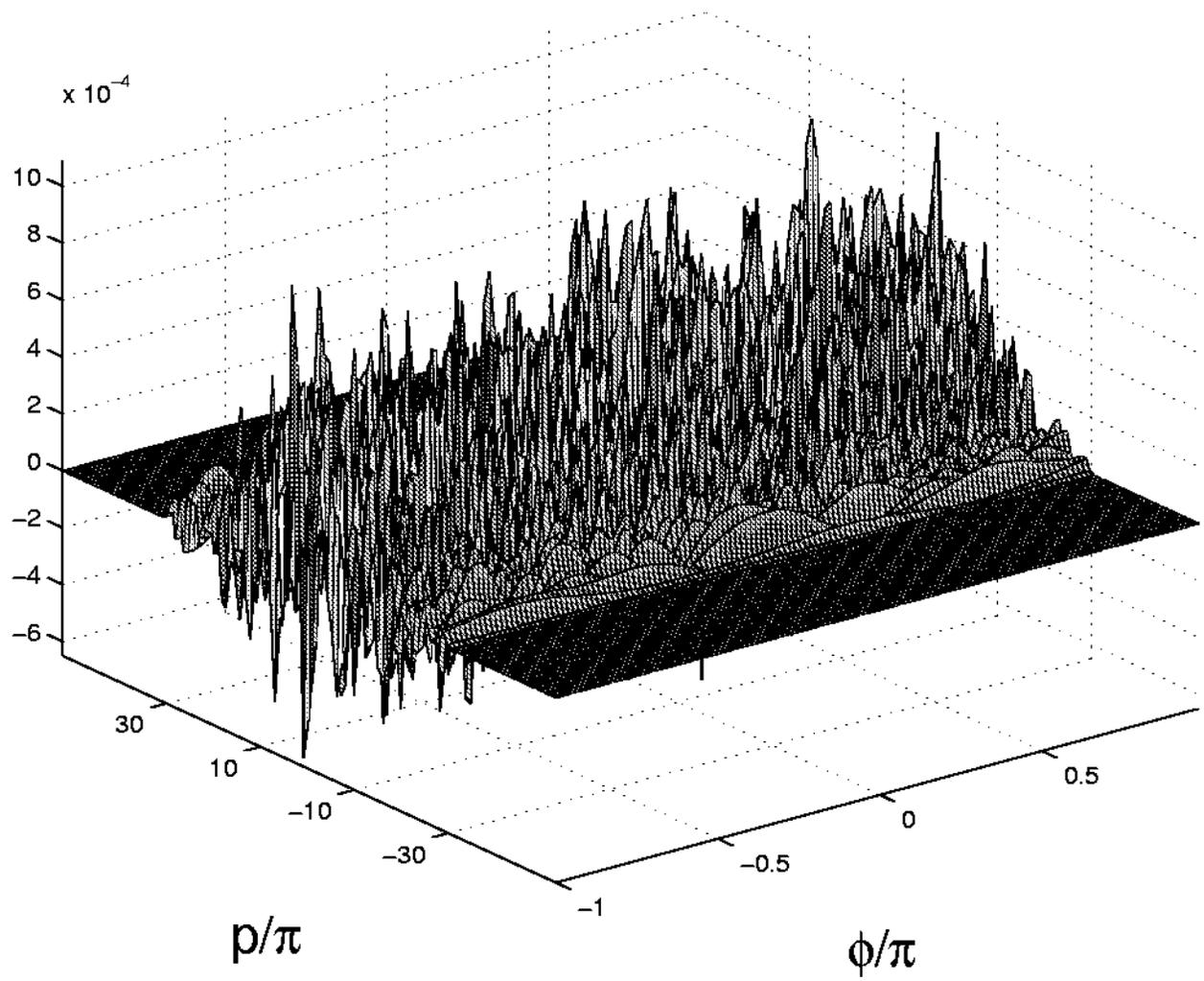
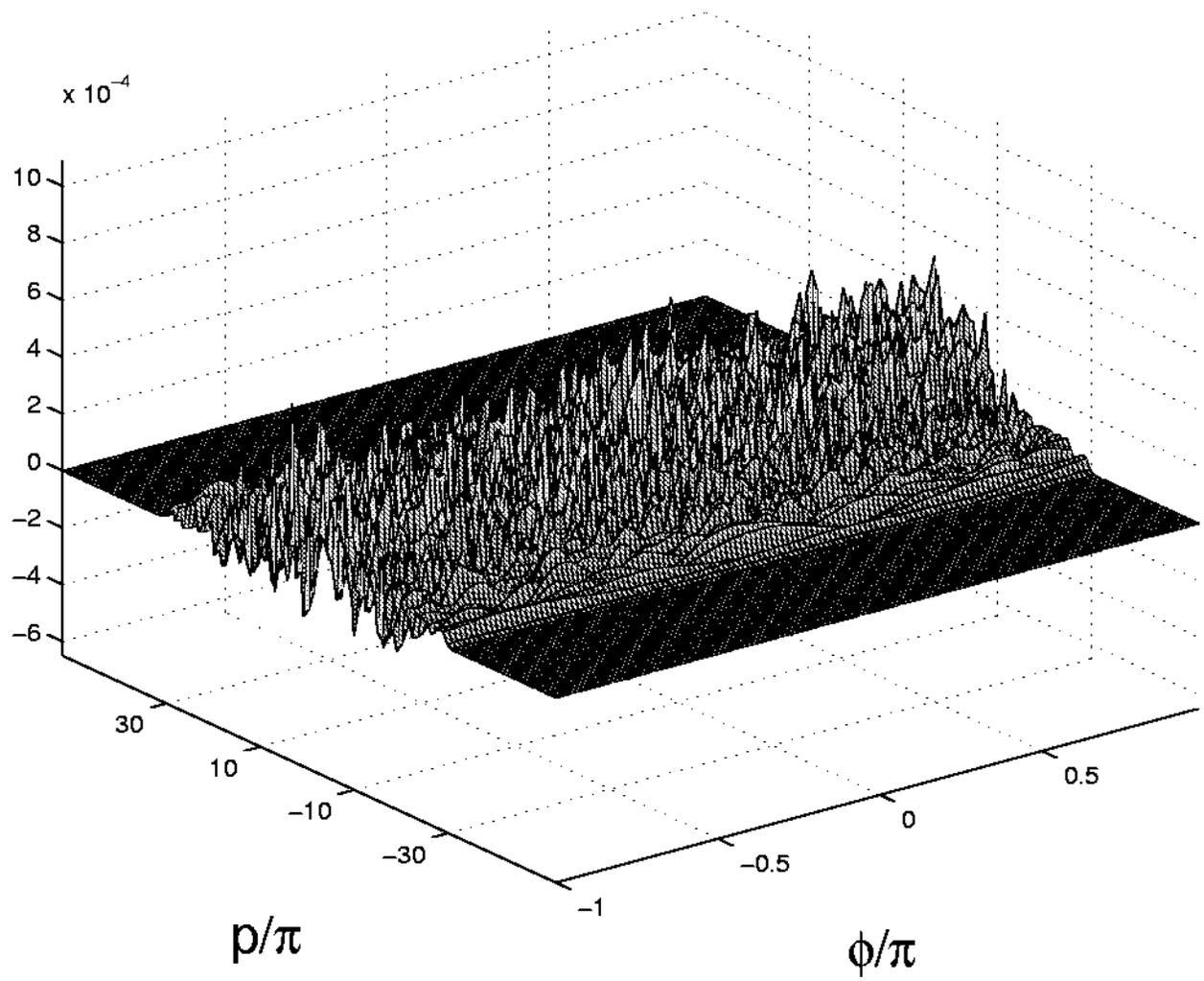
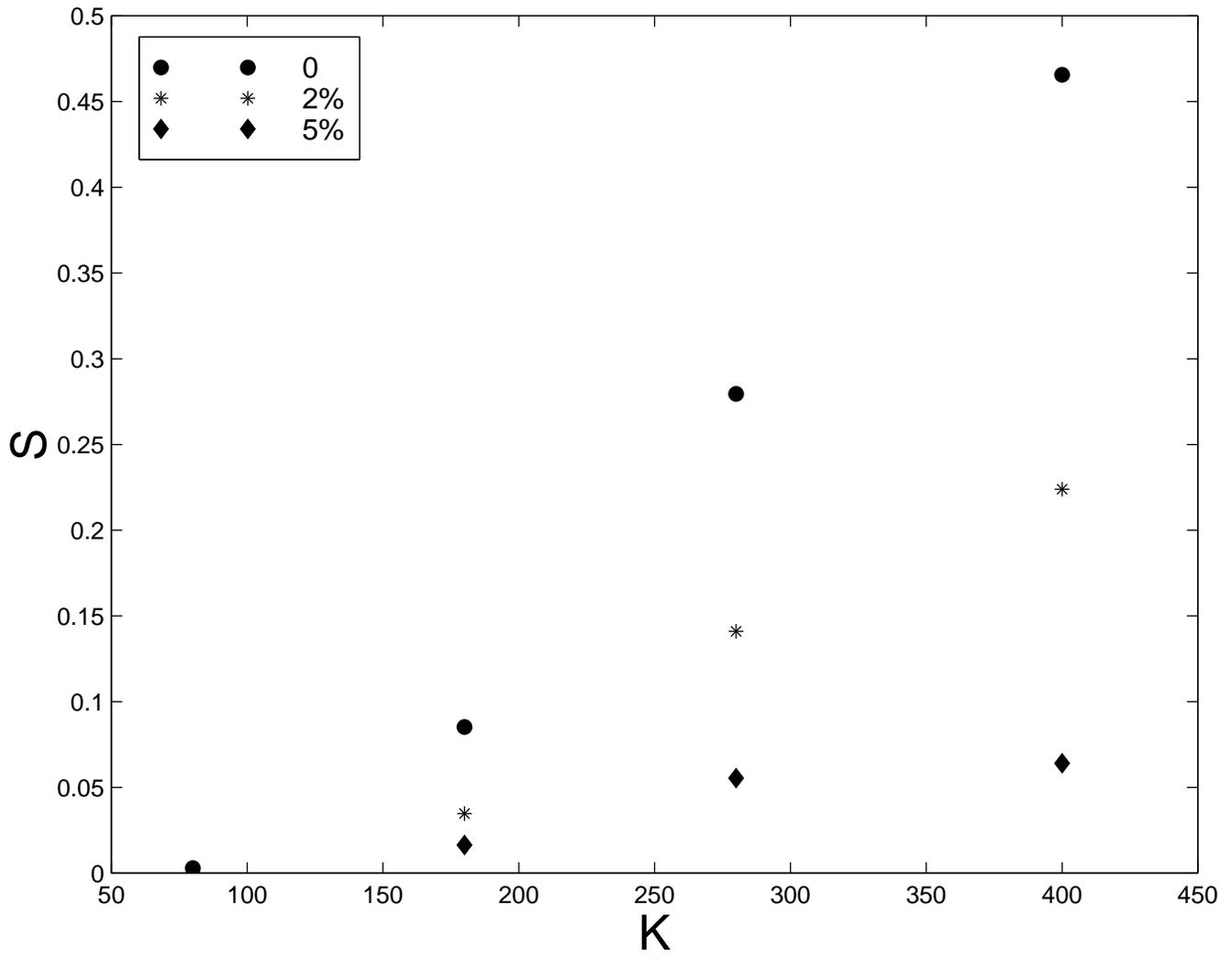
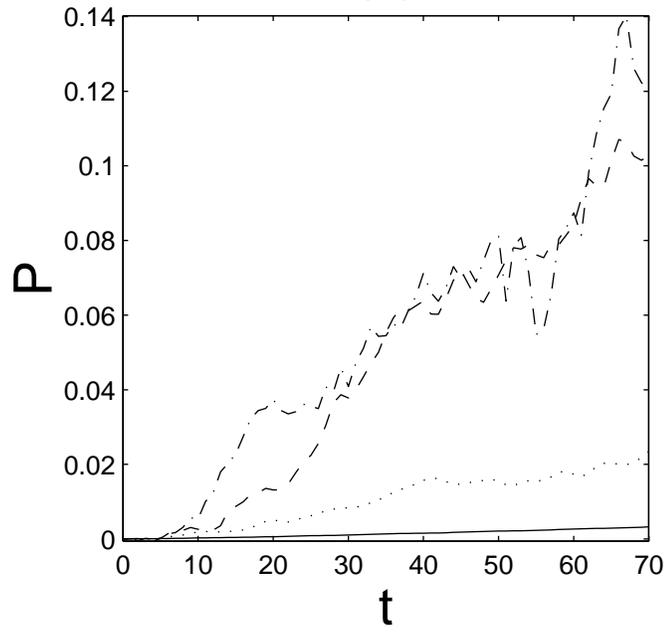
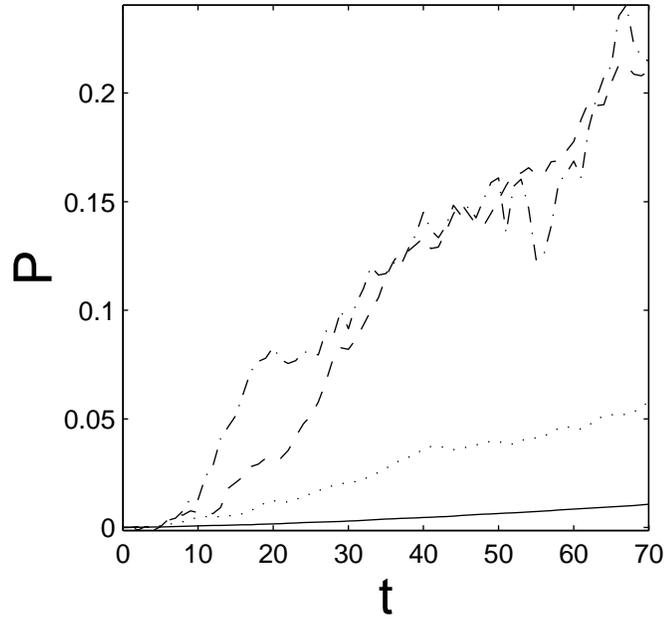
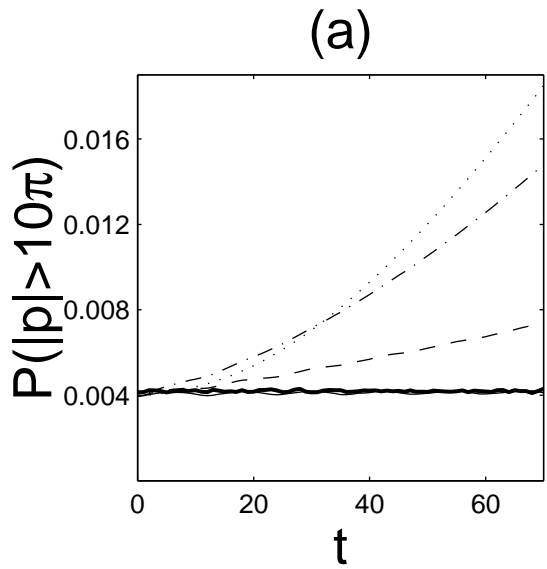
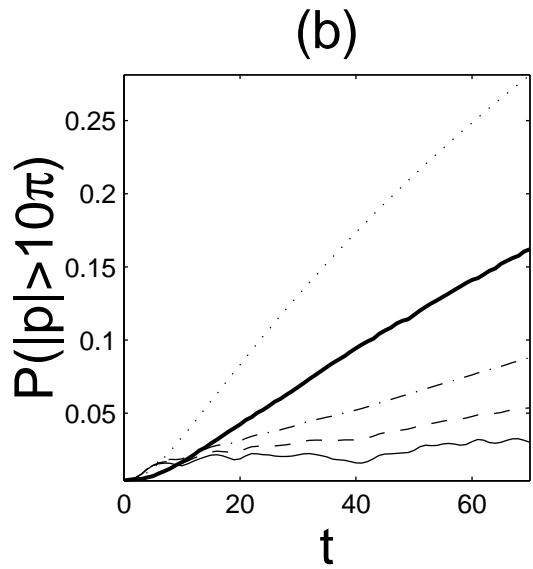
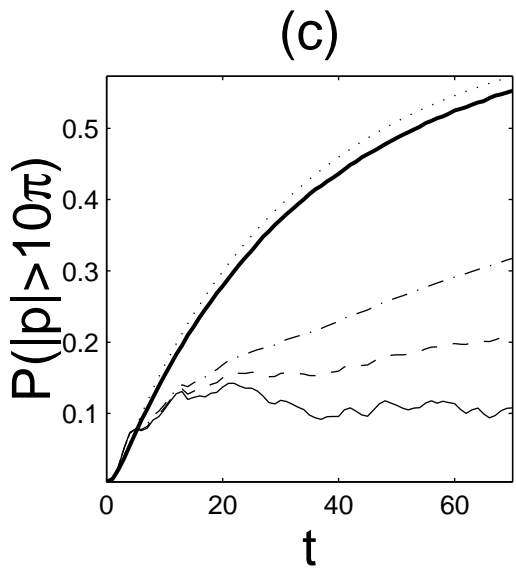
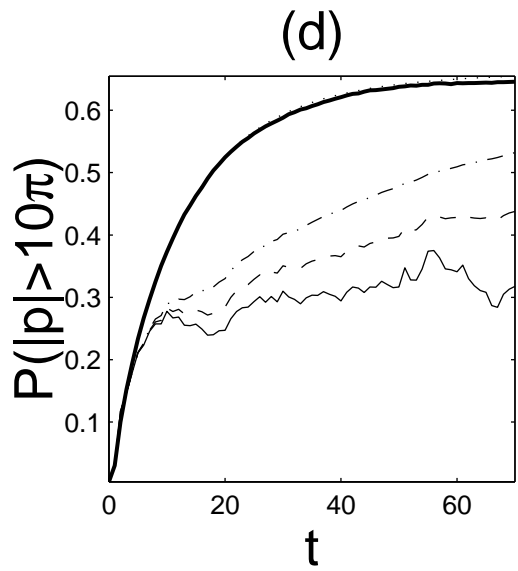
\end{document}